\RequirePackage{fix-cm}
\documentclass[smallextended]{svjour3}       
\smartqed  
\usepackage{graphicx}
\usepackage{mathptmx}      

\usepackage{latexsym}
\usepackage{geometry} 
\usepackage{amssymb}
\usepackage{amsmath,bm}    
\usepackage{wasysym}           	                  											

\begin{document}

\title{Investigation on a Doubly-Averaged Model for the Molniya Satellites Orbits}

\author{Tiziana Talu \and Elisa Maria Alessi\and Giacomo Tommei }


\institute{T. Talu \at Universit\`a di Pisa, Dipartimento di Matematica, Largo B. Pontecorvo 5, 56127 Pisa, Italy \\
	  \email{tiziana.talu92@gmail.com}
	\and 
E. M. Alessi \at IMATI-CNR, Istituto di Matematica Applicata e Tecnologie informatiche ``E. Magenes'', Consiglio Nazionale delle Ricerche, Via Alfonso Corti 12, 20133 Milano, Italy\\
\email{em.alessi@mi.imati.cnr.it}
\and
G. Tommei \at Universit\`a di Pisa, Dipartimento di Matematica, Largo B. Pontecorvo 5, 56127 Pisa, Italy\\
\email{giacomo.tommei@unipi.it}}

\date{Received: date / Accepted: date}

\maketitle

\begin{abstract}

The aim of this work is to investigate the lunisolar perturbations affecting the long-term dynamics of a Molniya satellite. Some numerical experiments on the doubly-averaged model, including the expansion of the lunisolar disturbing functions up to the third order, are carried out in order to detect the terms dominating the long-term evolution. The analysis focuses on the following significant indicators: the \textit{amplitude} of the harmonic coefficients, the \textit{periods} of the arguments involved and, in particular, the \textit{ratio} between the \textit{amplitudes}and the corresponding frequency. The results show that the second-order lunisolar perturbation gives the dominant contribution to the long-term dynamics. \\
The second part of this work aims to study the resonant regions associated to the dominant terms identified so far by using both the ideal resonance model and an alternative approach. The results obtained show when the standard method does not catch the main features of the dynamical structure of the resonant regions. Finally, the maximum overlapping region is identified in the proximity of the Molniya orbital environment.

\end{abstract}

\section{Introduction}
\label{intro}
On April 23, 1965 the first \textit{Molniya-1} spacecraft was launched by the former Soviet Union \cite{anselmo}. After that many others were set in orbit until 2004. These satellites were initially designed for Russian communication networks and their orbits form a class of special orbits around the Earth: the \textit{Molniya orbits}. 
The main dynamical features of Molniya orbits are: an eccentricity $e\geq 0.7$, an inclination $i\approx 63.4 \deg$ and an orbital period of approximately 12 hours.  Let us generically call \textit{Molniya satellite} a passive object orbiting along a Molniya type orbit. As a matter of fact, these satellites are no longer operational and thus they can be considered space debris.\\ The Russian territory to cover is enormous and located at a high latitude, thus an inclined stable apogee above the region of interest is needed. The inclination of the orbital plane is close to the \textit{critical inclination} value in such a way that the precession of the line of apsides induced by the oblateness of the Earth is cancelled out. It follows that the perigee and the apogee of the satellite remain almost frozen in time, according to the initial $\omega=270\hspace{0.1cm} \deg$ chosen due to the Russian latitude \cite{art}. Moreover, a Molniya satellite revolves two times around the Earth every day: in other words its orbital period and the Earth's rotation period are commensurable and this fact produces a \textit{tesseral resonance}. This is called \textit{mean motion resonances} in \cite{cinesimoln}, but it does not arise from a commensurability between mean motions; for this reason we prefer to use ``tesseral resonance'' throughout the discussion.\\
Because of its dynamical features, a Molniya satellite undergoes several perturbations. The low value of the altitude of the perigee, approximately $500\hspace{0.1cm}km$ \cite{wondermolniya}, gives a non-negligible atmospheric drag, which deeply affects the evolution of the semi-major axis. Besides, the satellite spends most of the time at high altitudes, hence the lunisolar effect plays a fundamental role on a timescale larger than the satellite orbital period.\\
In literature, the dynamics associated with Molniya orbits is faced following different perspectives. In \cite{cinesigeo} and \cite{DH} the perturbing effects of the geopotential are taken into account. In \cite{cinesigeo} they found that the value $\mathnormal{a}\approx 26554.3\hspace{0.1cm}km$ corresponds to the libration center of the $2:1$ tesseral resonance and the resonance width is $\Delta \mathnormal{a}\approx 38 \hspace{0.1cm}km$. Such geopotential-only model is not appropriate for the Molniya case, and the gravitational perturbation exerted by the Moon and by the Sun has been introduced in later works \cite{cinesimoln,DM}. Lunisolar effects are usually studied with a second order doubly-averaged model where the geocentric orbits of the third-bodies are circular. Under this assumption, the third order disturbing functions vanish, thanks to the analytical expressions of the eccentricity functions appearing in it \cite{frontiers}.\\
Molniya orbits are considered chaotic, sometimes the chaotic growth of the eccentricity leads to a dangerous low altitude of the perigee. To find the resonance location is useful to a explain chaotic behaviour; the \textit{Chirikov resonance overlapping criterion} states that when two or more critical arguments librate in the same region of the phase space a large-scale chaos may be expected, while the lack of overlapping between resonances usually guarantees the confinement of the motion \cite{morbidelli}. The web of secular lunisolar resonances in Medium Earth Orbit (MEO) region is usually explored approximating the slow frequencies of the satellites with the precession rate caused by the Earth oblateness \cite{meoreg,celletti}. Such approximation is generally both convenient and accurate enough but, as shown later in this  paper, it seems to be not appropriate to deal with the Molniya dynamics: the lunisolar contribution is not negligible especially for the dynamics of the argument of the perigee because of the critical inclination.\\
The purpose of this work is to investigate the long-term lunisolar perturbation affecting the Molniya dynamics and it will be structured as follows. In Sect.~\ref{sec:1} a brief overview of the theory behind the problem has been included while Sect.~\ref{sec:2} collects the results elaborated through the numerical investigation. We focus on a doubly-averaged model including the secular oblateness effect and the expansion of the lunar and solar disturbing functions up to the third order. By exploiting an analytical approach based on the Hamiltonian theory it is possible to identify the perturbing terms dominating the dynamics in the long-term. In this regard, the \textit{amplitudes} of the harmonic coefficients, the corresponding \textit{periods} and the \textit{ratio} between the \textit{amplitudes} and the corresponding frequency are estimated in the proximity of the Molniya orbital region (Sect.~\ref{sec:2.1}). Sect.~\ref{sec:phasespaceres} is dedicated to analyse the resonant dynamics associated to the main dominant terms identified in Sect.~\ref{sec:2.1}, assumed as isolated resonances. It will be shown from a theoretical and practical point of view (Sect.~\ref{sec:1resdynamics} and Sect.~\ref{sec:phasespaceres}, respectively) when the ideal resonance model does not produce an appropriate description of the resonances. As a matter of fact, resonant or near-resonant terms produce significant variations of the orbital elements on a long-term timescale. 

\section{Theoretical background}
\label{sec:1}
\subsection{Development of the dynamical model}
\label{sec:1model}
The analytical expressions of the perturbing forces can be easily found in literature but, in the majority of the cases, the developments are given in terms of Keplerian elements: the semi-major axis $\mathnormal{a}$, the eccentricity $e$, the inclination $i$, the argument of the perigee $\omega$, the longitude of the ascending node $\Omega$ and the mean anomaly $M$. Throughout the discussion we use the subscripts $\oplus$, $\leftmoon$ and $\odot$ to denote the parameters of the Earth, of the Moon or of the Sun, respectively. The satellite's elements will be denoted by no subscript. \\
Following \cite{celletti}, the orbital elements of the Sun with respect to the celestial equator are well approximated by linear functions of time, thus the solar disturbing function can be written as:
\begin{equation}\label{sunpot}
\begin{array}{ll}
\vspace{0.2cm}
\mathcal{R}_{\odot}= \sum_{l=2}^{\infty}  \sum_{m,p,q=0}^{l}  \sum_{j,r=-\infty}^{\infty}   \mu_{\odot} \big(\frac{\mathnormal{a}^l}{a_{\odot}^{l+1}}\big)\epsilon_m\frac{(l-m)!}{(l+m)!} F_{lmp}(\mathnormal{i})F_{lmq}(i_{\odot})H_{lpj}(\mathnormal{e})G_{lqr}(e_{\odot}) \times \\
\hspace{2cm}\times \cos\big[ (l-2p+j)M - (l-2q+r)M_{\odot} + (l-2p)\omega - (l-2q)\omega_{\odot} + m(\Omega-\Omega_{\odot}) \big]
\end{array}
\end{equation}
where $\mu_{\odot}$ is the gravitational parameter of the Sun and the Keplerian elements of both the satellite and the Sun are written with respect to the equatorial reference plane. $F_{lmp}$ and $F_{lmq}$ are Kaula's inclination functions, while $H_{lpj}$ and $G_{lqr}$ are Hansen coefficients.\\ The motion of the Moon around the Earth is quite perturbed by the Sun, hence the corresponding inclination, node and argument of the perigee with respect to the celestial equator evolve as nonlinear functions of time. However, if we adopt a mixed reference plane where the elements of the satellite are written with respect to the equatorial plane while the elements of the Moon are referred to the ecliptic, then $i_{\leftmoon}$ is approximately constant and $\omega_{\leftmoon}$ and $\Omega_{\leftmoon}$ are approximately linear functions of time \cite{celletti}. Because of the previous consideration, it is convenient to use the following disturbing function to better manipulate the lunar perturbation
 \begin{equation}\label{lunarpot}
\begin{array}{ll}
\vspace{0.2cm}
\mathcal{R}_{{\leftmoon}} = \sum_{l=2}^{\infty} \sum_{m,p,s,q=0}^l \sum_{j,r=-\infty}^{+\infty}(-1)^{m+s}(-1)^{k_1} \frac{\mu_{\leftmoon}\epsilon_m\epsilon_s}{2a_{\leftmoon}} \frac{(l-s)!}{(l+m)!}(\frac{a}{a_{\leftmoon}})^l  F_{lmp}(i)F_{lsq}(i_{\leftmoon})H_{lpj}(e)G_{lqr}(e_{\leftmoon})\times\\
\vspace{0.15cm}
\hspace{1cm}\times \Big\{ (-1)^{k_2}U_l^{m,-s}(\epsilon) \cos[(l-2p+j)M + (l-2q+r)M_{\leftmoon} + \\
\vspace{0.15cm}
\hspace{7cm} +(l-2p)\omega +(l-2q)\omega_{\leftmoon} +m\Omega + s(\Omega_{\leftmoon} - \frac{\pi}{2}) -y_s\pi] \\
\vspace{0.15cm}
\hspace{1cm} + (-1)^{k_3} U_l^{m,s}(\epsilon) \cos[(l-2p+j)M -(l-2q+r)M_{\leftmoon} +\\
\vspace{0.15cm}
\hspace{7cm} +(l-2p)\omega -(l-2q)\omega_{\leftmoon} +m\Omega -s(\Omega_{\leftmoon} - \frac{\pi}{2}) -y_s\pi]\Big\}.
\end{array}
\end{equation}
$\mu_{\leftmoon}$ is the gravitational parameter of the Moon, $\epsilon$ is the angle between the ecliptic and the equatorial plane, and $y_s=0$ for $s$ even while $y_s=\frac{1}{2}$ for $s$ odd. The analytical expansions of the Kaula's inclination functions and of $U_m^{l,\pm s}$, the Hansen coefficients and the coefficients $\epsilon_m$, $\epsilon_s$, $k_1$, $k_2$, $k_3$ can be found in \cite{celletti,kaula,laskar}. \\
We are interested in a model including the oblateness effect and the third-body perturbation up to the third order, that is including the harmonics in Eqs. $\eqref{sunpot}$ and $\eqref{lunarpot}$ with $l=2,3$. Usually, in order to investigate the long-term evolution a doubly-averaged model is used, thus, the disturbing potential considered is: 
\begin{equation}\label{pot}
	\mathcal{R}= \mathcal{\bar R}_{J_2} + \mathcal{\bar{\bar{R}}}_{\leftmoon} + \mathcal{\bar{\bar{R}}}_{\odot}
\end{equation}
The first term in Eq. $\eqref{pot}$ is the secular oblateness effect
\begin{equation}
	 \mathcal{\bar R}_{J_2} = \frac{1}{4}J_2 \frac{\mu_{\oplus}R_{\oplus}^2}{\mathnormal{a}^3(1-e^2)^{\frac{3}{2}}}\biggl(1-3\cos^2 i \biggr)
\end{equation} 
where $J_2$ is the second order zonal coefficient, $\mu_{\oplus}$ is the Earth's gravitational parameter and $R_{\oplus}$ represents the equatorial mean radius of the Earth. The terms $\mathcal{\bar{\bar{R}}}_{\leftmoon}$ and $\mathcal{\bar{\bar{R}}}_{\odot}$ in Eq. $\eqref{pot}$ are, respectively, the lunar and solar disturbing function averaged formerly over the mean anomaly of the satellite $M$ and then over the mean anomalies of the perturbing bodies. Since both $\mathcal{R}_{\leftmoon}$ and $\mathcal{R}_{\odot}$ are periodic functions of the angles, the doubly-averaged potentials $\mathcal{\bar{\bar{R}}}_{\leftmoon}$ and $\mathcal{\bar{\bar{R}}}_{\odot}$ are the collections of all the terms in Eqs. $\eqref{sunpot}$ and $\eqref{lunarpot}$, respectively, such that: 
\begin{equation}
	\left\{ \begin{array}{ll}
	l=2,3 \\
	l-2p+j = 0 \\
	l-2q+r =0 
	\end{array}\right.
\end{equation}
The averaging procedure is allowed whenever no mean motion resonance and no semi-secular lunisolar resonance occur, the latter arise from commensurabilities between the slow frequencies of the satellite and the mean anomalies of the Moon and the Sun. \\
In order to highlight the Hamiltonian structure of the problem a coordinate change is required to switch to the Dealunay canonical variables. In this way, the Hamiltonian describing the long-term lunisolar effect on a Molniya satellite is
\begin{equation}\label{hamiltonian}
	\mathcal{H}(L,G,H,\ell,g,h)= \mathnormal{H}_{kep}(L) + \mathcal{H}_{J_2}(G,H;L) + \mathcal{H}_{\leftmoon}(G,H,g,h;L) + \mathcal{H}_{\oplus}(G,H,g,h;L)
\end{equation}
where 
\begin{equation}
		\mathnormal{H}_{kep}=-\frac{\mu_{\oplus}^2}{2L^2} 
\end{equation}
and 
\begin{equation}
	\mathcal{H}_{J_2}= - \mathcal{\bar R}_{J_2}, \quad \mathcal{H}_{\leftmoon}= - \mathcal{\bar{\bar{R}}}_{\leftmoon}, \quad  \mathcal{H}_{\odot}= - \mathcal{\bar{\bar{R}}}_{\odot}
\end{equation}
written in terms of Delaunay variables.
It has to be pointed out that in Eq. $\eqref{lunarpot}$ the harmonic argument vanishes for $l=2$, $ p=1$ and $m,s=0$. Since the corresponding term in $\mathcal{H}_{\leftmoon}$ only depends on the actions $(L,G,H)$, we will call this special harmonic the \textit{lunar mean term}. As for the lunar case, for $l=2$, $ p=1$ and $m=0$, the solar harmonic argument in Eq. $\eqref{sunpot}$ disappears and thus we will refer to the corresponding harmonic term as the \textit{solar mean term}. 

\subsection{On the Hamiltonian dynamics}
\label{sec:1hamiltoniansystem}
With the use of the Delaunay variables, it is convenient to adopt a suitable notation. Let us denote: 
\begin{equation}\label{ham}
\begin{aligned}
	&\mathcal{H}_{\leftmoon}(G,H,g,h;L)= C^{\leftmoon}_{0} \mathcal{A}^{\leftmoon}_{0}(G,H;L) + \sum_{\alpha} C^{\leftmoon}_{\alpha} \mathcal{A}^{\leftmoon}_{\alpha}(G,H;L)\cos(\varphi^{\leftmoon}_{\alpha})\\	&\mathcal{H}_{\odot}(G,H,g,h;L)= C^{\odot}_{0} \mathcal{A}^{\odot}_{0}(G,H;L)+\sum_{\gamma} C^{\odot}_{\gamma} \mathcal{A}^{\odot}_{\gamma}(G,H;L)\cos(\phi^{\odot}_{\gamma})
	\end{aligned}
\end{equation}
where: 
\begin{itemize}
	\item $\alpha$ and $\gamma$ index the finite number of lunar and solar harmonics retained in the model, respectively;
	\item $C^{\leftmoon}_{\alpha} \mathcal{A}^{\leftmoon}_{\alpha}(G,H;L)$ is the $\alpha-$th lunar harmonic coefficient, the constant term $C^{\leftmoon}_{\alpha}$ includes the lunar orbital parameters. $ C^{\odot}_{\gamma} \mathcal{A}^{\odot}_{\gamma}(G,H;L)$ is the $\gamma-$th solar harmonic coefficient and $C^{\odot}_{\gamma}$ includes the solar orbital parameters. 
		\item $\alpha=0$ and $\gamma=0$ denote the mean terms, that is, $C^{\leftmoon}_{0} \mathcal{A}^{\leftmoon}_{0}(G,H;L)$ is the \textit{lunar mean term} and $C^{\odot}_{0} \mathcal{A}^{\odot}_{0}(G,H;L)$ is the \textit{solar mean term}.
	\item $\varphi^{\leftmoon}_{\alpha}$ is the $\alpha-$th lunar argument and $\phi^{\odot}_{\gamma}$ is the $\gamma-$th solar argument. 
\end{itemize}
Generally, both sine and cosine trigonometric functions appear in the development of the lunar disturbing function. However, in our case only cosine harmonics remain because of  the values that $s$ and $y_s$ in Eq. $\eqref{lunarpot}$ assume when $l=2,3$. \\
The mean anomaly is cyclic in the doubly-averaged Hamiltonian $\eqref{ham}$, hence the action $L$ is a first integral: it means that the semi-major axis is constant in the long-term. The dynamics of the $G,H,g$ and $h$ is given by the following Hamilton equations: 
 \begin{equation}\label{hamsys}
\left\{\begin{aligned}
\vspace{0.2cm}
\dot G &= \sum_{\alpha} \biggl[C_{\alpha}^{\leftmoon}\mathcal{A}_{\alpha}^{\leftmoon}(G,H;L)\biggr]\frac{\partial \varphi^{\leftmoon}_{\alpha}}{\partial g}\sin(\varphi_{\alpha}^{\leftmoon})+ \sum_{\gamma}\biggl[C_{\gamma}^{\odot}\mathcal{A}_{\gamma}^{\odot}(G,H;L)\biggr]\frac{\partial \phi^{\odot}_{\gamma}}{\partial g}\sin(\phi_{\gamma}^{\odot}) \\
\vspace{0.2cm}
\dot H &= \sum_{\alpha} \biggl[C_{\alpha}^{\leftmoon}\mathcal{A}_{\alpha}^{\leftmoon}(G,H;L)\biggr]\frac{\partial \varphi^{\leftmoon}_{\alpha}}{\partial h}\sin(\varphi_{\alpha}^{\leftmoon})+ \sum_{\gamma}\biggl[C_{\gamma}^{\odot}\mathcal{A}_{\gamma}^{\odot}(G,H;L)\biggr]\frac{\partial \phi^{\odot}_{\gamma}}{\partial h}\sin(\phi_{\gamma}^{\odot}) \\
\vspace{0.2cm}
\dot{g}&= \frac{\partial \mathcal{H}_{J_2}}{\partial G}(G,H;L)+C^{\leftmoon}_{0}\frac{\partial \mathcal{A}^{\leftmoon}_{0}(G,H;L)}{\partial G} +C^{\odot}_{0}\frac{\partial \mathcal{A}^{\odot}_{0}(G,H;L)}{\partial G} + \\
&\hspace{4cm} +\sum_{\alpha} \biggl[C_{\alpha}^{\leftmoon}\frac{\partial\mathcal{A}_{\alpha}^{\leftmoon}}{\partial G}(G,H;L)\biggr] \cos(\varphi_{\alpha}^{\leftmoon})+ \sum_{\gamma}\biggl[C_{\gamma}^{\odot}\frac{\partial\mathcal{A}_{\gamma}^{\odot}}{\partial G}(G,H;L)\biggr]\cos(\phi_{\gamma}^{\odot}) \\
\dot h &=  \frac{\partial \mathcal{H}_{J_2}}{\partial H}(G,H;L) + C^{\leftmoon}_{0}\frac{\partial \mathcal{A}^{\leftmoon}_{0}(G,H;L)}{\partial H} +C^{\odot}_{0}\frac{\partial \mathcal{A}^{\odot}_{0}(G,H;L)}{\partial H} + \\
&\hspace{4cm}+\sum_{\alpha}\biggl[ C_{\alpha}^{\leftmoon}\frac{\partial \mathcal{A}_{\alpha}^{\leftmoon}}{\partial H}(G,H;L)\biggr]\cos(\varphi_{\alpha}^{\leftmoon})+ \sum_{\gamma}\biggl[C_{\gamma}^{\odot}\frac{\partial \mathcal{A}_{\gamma}^{\odot}}{\partial H}(G,H;L)\biggr]\cos(\phi_{\gamma}^{\odot}) \\
\end{aligned}\right.
\end{equation}

The oblateness of the Earth does not produce any effect on the actions $G$ and $H$, but it causes a precession, or regression, of $g$ and $h$ which is usually used to approximate the evolution of the angles, as already mentioned before. From the last two equations of the system $\eqref{hamsys}$ we get that the angles undergo secular drifts, caused by the oblateness and by the lunar and solar mean terms, and periodic effects, given by integrating the oscillating terms whose amplitude is proportional to the partial derivatives of the harmonic coefficients. Since the \textit{Laplace radius} is around $7.7\hspace{0.1cm} R_{\oplus}$ \cite{tremaine}, that is the geocentric distance for which the order of magnitude of the precession caused by the lunisolar perturbation is equivalent to the one caused by the Earth oblateness, the following approximation 
\begin{equation}
	\left\{\begin{array}{ll}
	\vspace{0.12cm}
	 \dot g \approx \frac{\partial \mathcal{H}_{J_2}}{\partial G} \\
	 \dot h \approx \frac{\partial \mathcal{H}_{J_2}}{\partial H}\\
	 \end{array}\right.
\end{equation}
is \textit{usually} both convenient and accurate enough. However, in the particular case of the Molniya dynamics, the orbits are critical inclined, thus the third-body perturbation might not be necessarily negligible at least for $\dot g$, as confirmed by numerical experiments that will be presented in Sect.~\ref{sec:2.1}. \\
The first two equations of the system $\eqref{hamsys}$ suggest that the larger the harmonic coefficient the deeper the resulting fluctuations in $\dot G$ and $\dot H$. A quantity that particularly matters concerning the evolution over time of $G$ and $H$ is the \textit{ratio} between the amplitude of the harmonic coefficients and the corresponding frequency. Let us consider a first approximation of the system $\eqref{hamsys}$ where: 
\begin{itemize}
\item the actions are assumed constants 
\begin{equation}
	\mathcal{A}_{\alpha}^{\leftmoon}(G,H;L)= \mathcal{A}_{\alpha}^{\leftmoon}, \quad {\mathcal{A}}_{\gamma}^{\odot}(G,H;L)={\mathcal{A}}_{\gamma}^{\odot}
\end{equation}
\item the angles evolve linearly in time
\begin{equation}\label{linearangle}
\varphi_{\alpha}^{\leftmoon}(t)= \varphi_{\alpha,0}^{\leftmoon}+ \dot{\varphi}_{\alpha}^{\leftmoon}t, \quad 	\phi_{\gamma}^{\odot}(t)= \phi_{\gamma,0}^{\odot}+\dot{\phi}_{\gamma}^{\odot}t\\
\end{equation}
being $\varphi_{\alpha,0}^{\leftmoon}$ and $\phi_{\gamma,0}^{\odot}$ generic initial conditions and $\dot{\varphi}_{\alpha}^{\leftmoon}$ and $\dot{\phi}_{\gamma}^{\odot}$ constants. 
\end{itemize}
In this way, the system $\eqref{hamsys}$ is approximated by 
\begin{equation}\label{oscillatingactionsc}
\left\{\begin{aligned}
\vspace{0.2cm}
\dot G &= \sum_{\alpha}	C_{\alpha}^{\leftmoon}\mathcal{A}_{\alpha}^{\leftmoon}\frac{\partial \varphi^{\leftmoon}_{\alpha}}{\partial g} \sin(\varphi_{\alpha,0}^{\leftmoon}+ \dot{\varphi}_{\alpha}^{\leftmoon}t)+ \sum_{\gamma}C_{\gamma}^{\odot}\mathcal{A}_{\gamma}^{\odot} \frac{\partial \phi^{\odot}_{\gamma}}{\partial g} \sin( \phi_{\gamma,0}^{\odot}+\dot{\phi}_{\gamma}^{\odot}t)\\
\dot H &= \sum_{\alpha}C_{\alpha}^{\leftmoon}\mathcal{A}_{\alpha}^{\leftmoon}\frac{\partial \varphi^{\leftmoon}_{\alpha}}{\partial h}\sin(\varphi_{\alpha,0}^{\leftmoon}+ \dot{\varphi}_{\alpha}^{\leftmoon}t)+ \sum_{\gamma}C_{\gamma}^{\odot}\mathcal{A}_{\gamma}^{\odot} \frac{\partial \phi^{\odot}_{\gamma}}{\partial h}\sin( \phi_{\gamma,0}^{\odot}+\dot{\phi}_{\gamma}^{\odot}t) \\
\end{aligned}\right.
\end{equation}

Now, it is easy to integrate the system $\eqref{oscillatingactionsc}$ on a timespan $[0,T]$, because the indices in the summations are in a finite number. If $\Delta G=G(T)-G(0)$ and $\Delta H=H(T)-H(0)$, then:

\begin{equation}
\begin{aligned}
&\Delta G = \sum_{\alpha}\frac{C_{\alpha}^{\leftmoon}\mathcal{A}_{\alpha}^{\leftmoon}}{\dot \varphi_{\alpha}^{\leftmoon}}\frac{\partial \varphi^{\leftmoon}_{\alpha}}{\partial g} \biggl[\cos(\varphi_{\alpha,0}^{\leftmoon}) -\cos(\varphi_{\alpha}^{\leftmoon}(T))\biggr]+ \sum_{\gamma}\frac{C_{\gamma}^{\odot}\mathcal{A}_{\gamma}^{\odot} }{\dot \phi_{\gamma}^{\odot}} \frac{\partial \phi^{\odot}_{\gamma}}{\partial g} \biggl[\cos(\phi_{\gamma,0}^{\odot}) -\cos(\phi_{\gamma}^{\odot}(T))\biggr] \\
&\Delta H = \sum_{\alpha}\frac{C_{\alpha}^{\leftmoon}\mathcal{A}_{\alpha}^{\leftmoon}}{\dot \varphi_{\alpha}^{\leftmoon}} \frac{\partial \varphi^{\leftmoon}_{\alpha}}{\partial h} \biggl[\cos(\varphi_{\alpha,0}^{\leftmoon}) -\cos(\varphi_{\alpha}^{\leftmoon}(T))\biggr] + \sum_{\gamma}\frac{C_{\gamma}^{\odot}\mathcal{A}_{\gamma}^{\odot}}{\dot \phi_{\gamma}^{\odot}} \frac{\partial \phi^{\odot}_{\gamma}}{\partial h}\biggl[\cos(\phi_{\gamma,0}^{\odot}) -\cos(\phi_{\gamma}^{\odot}(T))\biggr]\\
\end{aligned}
\end{equation} 
Under this approximation, $G$ and $H$ undergo periodic or secular effects depending on the ratio between the amplitudes of the harmonic coefficients and the corresponding frequency.
The larger the ratio, the deeper the long-term effects are. 
As a matter of fact, the near-resonant terms produce \textit{small divisors} and cause a substantial variations over time. The behaviour of the approximated solutions allow us to identify the dominant perturbing terms and the negligible ones also for the not-approximated orbit. In \cite{murray} it can be found similar results concerning the main mean motion resonances in the main asteroid belt.  \\


\subsection{The resonant dynamics}
\label{sec:1resdynamics}
\begin{figure}
	\centering
	\begin{minipage}{0.4\textwidth}
		\includegraphics[width=0.85\textwidth]{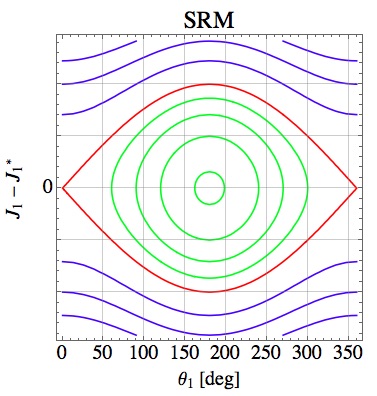}\\
	\end{minipage}%
	\begin{minipage}{0.4\textwidth}
		\includegraphics[width=0.91\textwidth]{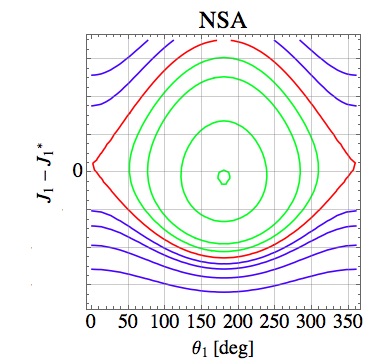}\\
	\end{minipage}
	\caption{On the left, the typical phase portrait of a pendulum-like dynamics. The libration region is symmetric with respect the axis $J_1=J_*$. On the right: phase portrait of an asymmetrical case. The plot is centered in the exact resonance, in order to appreciate that the unstable equilibrium is located above the line $J_1=J_*$ and the stable one is below. The resonant region on the right stretches upwards and the symmetry found on the left completely disappears.}
	\label{fig:penddin}   
\end{figure}

A non-autonomous dynamical system can be converted in an autonomous one by adding one dimension to the phase space. Therefore, without loss of generality, in what follows we assume to have an autonomous $N-$degree of freedom nearly-integrable Hamiltonian system. Referring to the classical theory presented in \cite{morbidelli}, let us take into account as a concrete example, useful to our purpose, the resonant Hamiltonian:
\begin{equation}\label{resham}
	\mathcal{H}_{res}(\mathbf{I}, \mathbf{\psi})= \mathcal{H}_0(\mathbf{I})+ \varepsilon f(\mathbf{I})\cos(\mathbf{k}\cdot \bm{\psi})
\end{equation}
where $\varepsilon$ is the small parameter, $\mathbf{I}\in \mathbb{R}^3$ and $\bm{\psi}\in \mathbb{T}^3$ are the action-angle variables\footnote{We use the following notation to denote the components of a generic vector $\mathbf{v}\in \mathbb{R}^3$: $v_i=\mathbf{e}_i^T\cdot \mathbf{v}$ where $\{\mathbf{e}_1,\mathbf{e}_2,\mathbf{e}_3\}$ is the canonical basis of $\mathbb{R}^3$} for the unperturbed Hamiltonian $\mathcal{H}_0$, and $\mathbf{k}\cdot\bm{\dot{\psi}}=0$ in some region of the phase space. The Hamilton equations arising from $\mathcal{H}_{res}$ are:
\begin{equation}
	\left\{ \begin{array}{ll}
	\dot{\mathbf{I}}= \varepsilon \mathbf{k}\cdot f(\mathbf{I})\sin (\mathbf{k}\cdot \bm{\psi})\\
	\dot{\bm{\psi}}= \frac{\partial \mathcal{H}_0}{\partial \mathbf{I}}(\mathbf{I}) + \varepsilon\frac{\partial f}{\partial \mathbf{I}}(\mathbf{I})\cos(\mathbf{k}\cdot \bm{\psi})	\end{array}\right.
\end{equation}
where $\frac{\partial \mathcal{H}_0}{\partial \mathbf{I}}(\mathbf{I})$ is the vector of the \textit{main frequencies}. From the classical theory, by \textit{resonance} is meant a commensurability between the main frequencies for some value of $\mathbf{I}=\mathbf{I}^*$, in this case: 
\begin{equation}\label{exres}
	\mathbf{k}\cdot \frac{\partial \mathcal{H}_0}{\partial \mathbf{I}}(\mathbf{I}^*) =0 
\end{equation}
In this work we need to make some distinctions.  We refer to the relation in Eq. $\eqref{exres}$ calling it the \textit{exact resonance}, while we talk about \textit{real resonance}, or simply \textit{resonance}, by referring to the following relation: 
\begin{equation}\label{realres}
	\mathbf{k}\cdot \dot{\bm{\psi}}(\mathbf{I})= \mathbf{k}\cdot \frac{\partial \mathcal{H}_0}{\partial \mathbf{I}}(\mathbf{I})  + \varepsilon\mathbf{k}\cdot \frac{\partial f}{\partial \mathbf{I}}(\mathbf{I})\cos(\mathbf{k}\cdot \bm{\psi}) =0 
\end{equation}
If the perturbation is sufficiently small with respect to the unperturbed dynamics, then $\mathbf{k}\cdot \dot{\bm{\psi}}(\mathbf{I}^*) \approx 0$ and the \textit{exact resonance} may well-approximate the \textit{real resonance} at least up to the first order in $\varepsilon$. 
There always exists a canonical transformation $\Phi$ such that the \textit{critical argument} $\mathbf{k}\cdot \bm{\psi}$ is a new angle, that is: 
\begin{equation}\label{resvar}
	(\mathbf{I}, \bm{\psi})\stackrel{\Phi}{\mapsto}(\mathbf{J}, \bm{\theta}), \quad \theta_1=\mathbf{e_1}^T\cdot \bm{\theta}= \mathbf{k}\cdot \bm{\psi}.
\end{equation}
After performing a coordinate change $\Phi$, the new Hamiltonian 
\begin{equation}
	\mathcal{H}'_{res}(J_1,\theta_1)= \mathcal{H}'_0(\mathbf{J}) + \varepsilon f'(\mathbf{J})\cos\theta_1
\end{equation}
describes a two dimensional motion taking place along the level curves $J_2=J_2^*$ and $J_3=J_3^{*}$ in the $(J_1, \theta_1)$ plane, where: $J_1$ is the action conjugate to the critical angle $\theta_1$ and $\bm{J}^*=\Phi(\mathbf{I}^*)$. \\
According to the Standard Resonance Model (SRM) \cite{morbidelli}, the Hamiltonian $\mathcal{H}'_{res}$ can be developed in Taylor series of $J_1$ around $J_1^{*}$ up to the second order. If we neglect the perturbing terms of the first order in $(J_1-J_1^{*})$ and higher, we obtain the so-called pendulum-like Hamiltonian 
\begin{equation}
	\mathcal{H}''_{res}(J_1,\theta_1)= \frac{\beta}{2}(J_1-J^{*}_1)^2 + c\cos\theta_1; \quad \beta= \frac{\partial^2\mathcal{H}'_0}{\partial J^{2}_1}(\mathbf{J}^{*}), \quad c= \varepsilon f'(\mathbf{J}^{*})
\end{equation}

describing a pendulum-like dynamics in the proximity of the exact resonance (Fig.~\ref{fig:penddin} on the left). Following \cite{morbidelli}, the resonant region is the libration region around the stable equilibria and its \textit{maximum libration width} measured at the apex of the separatrix is given by
\begin{equation}\label{maxlibwidth}
	|J_1-J_1^{*}|\leq 2\sqrt{\biggl|\frac{c}{\beta}\biggr|}
\end{equation}
If there are two or more resonance, then we can separately study the dynamics corresponding to each one making the assumption that they are isolated. The resulting motion is pendulum-like with appropriate coordinate change for every single resonance and the pendulum-like model may give a well approximation as long as the libration regions remain isolated. If any resonances overlap occurs, then the pendulum-like model breaks down. The separatrices of different resonances are connected if two or more resonances overlap, therefore an initial condition in this region may produce jumps from one libration region to one other showing chaotic diffusion \cite{morbidelli}. \\
Another scenario in which the classical approach does not provide a reliable description of the real resonant dynamics occurs when the exact resonance is not a well-approximation of the real resonance. The real equilibria arising from the suitable Hamiltonian $\mathcal{H}'_{res}$ are solutions of: 
\begin{equation}\label{realeq}
	\left\{ \begin{array}{ll}
	\dot{J}_1= \varepsilon f'(\mathbf{J})\sin \theta_1 =0\\
	\dot{\theta}_1=\frac{\partial \mathcal{H}'_0}{\partial J_1}(\mathbf{J}) + \varepsilon \frac{\partial f'}{\partial J_1}(\mathbf{J}) \cos \theta_1 =0
	\end{array}\right.
\end{equation}
As in the pendulum case, $\dot{J}_1=0 $ implies:
\begin{equation}\label{j1eq}
\theta_1= n\pi, \hspace{0.3cm} n\in \mathbb{Z}
\end{equation}
By replacing the solution $\eqref{j1eq}$ in the second equation of $\eqref{realeq}$, then this last splits in two different equations: 
\begin{equation}\label{realeq1}
\left\{\begin{array}{ll}
\vspace{0.2cm}
\frac{\partial \mathcal{H}'_0}{\partial J_1}(\mathbf{J}) +\varepsilon\frac{\partial f'}{\partial J_1}(\mathbf{J})=0\\
\frac{\partial \mathcal{H}'_0}{\partial J_1}(\mathbf{J}) -\varepsilon\frac{\partial f'}{\partial J_1}(\mathbf{J})=0
\end{array}\right.
\end{equation}
Hence, solutions of $\eqref{realeq1}$ are not necessarily the same, the stronger the perturbing effects, controlled by $\varepsilon$, with respect to $\frac{\partial \mathcal{H}'_0}{\partial J_1}(\mathbf{J})$ the more the solutions of the system $\eqref{realeq1}$ are separated in the phase space. In such case, the Taylor approximation, which characterizes the classical approach, fails to catch a \textit{deep asymmetry}. Fig.~\ref{fig:penddin} on the right displays the phase portrait of a deep asymmetric case, where the stable and the unstable equilibria do not lie on the same line. \\
Let us assume that $(J_{1s},\theta_{1s})$ is the stable equilibrium of the system $\eqref{realeq}$ and $(J_{1u},\theta_{1u})$ is the unstable one, such that $J_{1s}\not = J_{1u}$. 
The maximum and the minimum value of $J_1$, $J_{max}$ and $J_{min}$ respectively, at the edge between the libration region and the separatrices are solution of: 
\begin{equation}\label{width}
	\mathcal{H}'_{res}(J_1,\theta_{1s})=\mathcal{H}'_{res}(J_{1u},\theta_{1u})
\end{equation}
The relation $\eqref{width}$ means that $J_{max}$ and $J_{min}$ are the intersections between $\theta_1=\theta_{1s}$ and the contour line, on the phase portrait, at the level $\mathcal{H}'_{res}(J_{1u},\theta_{1u})$, that is the contour line identifying the upper and the lower separatrix.\\
Actually, the maximum libration width in Eq. $\eqref{maxlibwidth}$ in the SRM is obtained following the same idea, but for a \textit{symmetric} situation produced by a Hamiltonian developed in Taylor series for which $J_{1s}= J_{1u}= J_1^{*}$. Hence it well describes a symmetric case where $|J_{1s}- J_{1u}|$ is null or sufficiently small. Conversely, the relation $\eqref{width}$, obtained with a not-standard approach (NSA), gives a more reliable range $[J_{min}, J_{max}]$ of the resonant region in a deep asymmetrical case.

\section{Numerical Results}
\label{sec:2}
In this section we show the numerical experiments on the doubly-averaged model of Eq. $\eqref{hamiltonian}$. The following values are used in what follows:
\begin{equation*}
\mathnormal{a}_{moln}= 26554.3\hspace{0.1cm} km, \quad
e_{moln}= 0.72, \quad 
i_{moln}= 63.43\hspace{0.1cm} \deg
\end{equation*}
and
\begin{equation*}
L_{moln}= \sqrt{\mu_{\oplus}\mathnormal{a}_{moln}}\hspace{0.1cm}\frac{km^2}{s}, \quad 
G_{moln}= L_{moln}\sqrt{1-e^2_{moln}} \hspace{0.1cm}\frac{km^2}{s}, \quad
H_{moln}= G_{moln}\cos i_{moln} \hspace{0.1cm}\frac{km^2}{s}
\end{equation*}
We refer to the above parameters as the \textit{Molniya parameters}. For the sake of consistency we also use the Delaunay angles for both the Moon and the Sun:
\begin{equation}
 \left\{ \begin{array}{ll}
	\omega_{\leftmoon}=g_{\leftmoon}\\
	\Omega_{\leftmoon}= h_{\leftmoon}
	\end{array}\right., \quad \left\{ \begin{array}{ll}
      \omega_{\odot}=g_{\odot}\\
      \Omega_{\odot}=h_{\odot}
	\end{array}\right.
\end{equation} 
Important results will be translated in terms of Keplerian elements in order to be more understandable. 

\subsection{The dominant terms in the long-term dynamics}
\label{sec:2.1}
According to the theoretical considerations exposed in Sect.~\ref{sec:1hamiltoniansystem}, we evaluate the \textit{amplitudes} of the harmonic coefficients (see Tabs.~\ref{tab:ampsole} and~\ref{tab:ampluna}), their partial derivatives with respect to the actions (see Tab. ~\ref{tab:deramp}), the \textit{periods} of the harmonic arguments (see Tabs.~\ref{tab:periodi1} and~\ref{tab:periodi2}) at the Molniya parameters. Since the functions involved are properly regular, the results provide an accurate estimate of the entity of the perturbing terms affecting a satellite in Molniya regime.\\
\begin{table}[h]
\caption{Largest amplitudes $[\frac{km^2}{s^2}]$, in absolute value, of the solar harmonic coefficients, together with the corresponding  argument. The values are computed by evaluating the harmonic coefficients at the Molniya parameters.}
\label{tab:ampsole}  
\centering
\begin{tabular}{ll}
\hline\noalign{\smallskip}
$\phi_{\gamma}^{\odot}$, $l=2$& $|C_{\gamma}^{\odot}\mathcal{A}_{\gamma}^{\odot}(L_{moln},G_{moln},H_{moln})|$\\
\noalign{\smallskip}\hline\noalign{\smallskip}
$2g$& $8.291\times10^{-6}$ \\
$2g+(h-h_{\odot})$& $6.420\times10^{-6}$  \\
$h-h_{\odot}$ & $5.442\times10^{-6}$\\
$ 2g-(h-h_{\odot})$& $2.451\times10^{-6} $ \\
Mean Term & $1.894\times 10^{-6}$\\
$ 2(h-h_{\odot})$ &  $ 1.179\times10^{-6}$ \\
$ 2g+ 2(h-h_{\odot}) $& $1.126\times10^{-6}$ \\
$ 2g-2(h-h_{\odot})$ & $1.642\times 10^{-7}$\\
\noalign{\smallskip}\hline
\end{tabular}
\end{table}
\begin{table}
	\caption{Largest amplitudes $[\frac{km^2}{s^2}]$, in absolute value, of the lunar harmonic coefficients, together with the corresponding  argument.}
	\label{tab:ampluna}  
	\centering
	\begin{tabular}{lllll}
       \cline{1-2} \cline{4-5}\noalign{\smallskip}
		$\varphi_{\alpha}^{\leftmoon}$, $l=2$& $|C_{\alpha}^{\leftmoon}\mathcal{A}_{\alpha}^{\leftmoon}(L_{moln},G_{moln},H_{moln})|$ &&$\varphi_{\alpha}^{\leftmoon}$,  $l=2$& $|C_{\alpha}^{\leftmoon}\mathcal{A}_{\alpha}^{\leftmoon}(L_{moln},G_{moln},H_{moln})|$  \\
			\noalign{\smallskip}\cline{1-2} \cline{4-5}\noalign{\smallskip}
		$2g$ & $1.791 \times10^{-5}$& & $2g+2h+h_{\leftmoon} $ & $ 4.562\times10^{-8}$ \\
		$2g +h$ &  $1.387 \times10^{-5} $&   & $2g-h+2h_{\leftmoon}$& $ 2.252\times10^{-8}$  \\
		$h$    & $1.176 \times10^{-5}$ & &$2g-2h+2h_{\leftmoon} $  & $ 1.677\times10^{-8}$ \\
		$2g -h$ & $ 5.296\times10^{-6}$ && $2g-2h_{\leftmoon}$ &$1.134\times10^{-8}$  \\
		$ 2g+h-h_{\leftmoon} $& $2.750\times10^{-6} $&& $2g+2h_{\leftmoon}$ & $1.134\times10^{-8} $  \\
		$2h$&$2.548\times10^{-6}$ & & Mean Term & $4.092\times10^{-6}$\\
		\noalign{\smallskip}\cline{4-5}
		$ 2g+2h $ & $2.432\times10^{-6}$&&$\varphi_{\alpha}^{\leftmoon}$,  $l=3$ & $|C_{\alpha}^{\leftmoon}\mathcal{A}_{\alpha}^{\leftmoon}(L_{moln},G_{moln},H_{moln})|$ \\
		\noalign{\smallskip}\cline{4-5}
		$ h-h_{\leftmoon}$ & $2.331 \times10^{-6} $ &&&\\
		$2g+ h_{\leftmoon} $& $1.162\times10^{-6}$&& $ 3g - g_{\leftmoon}+h - h_{\leftmoon} $  & $5.922\times10^{-8}$ \\
		$2g-h_{\leftmoon}$ &  $ 1.162\times10^{-6}$  &&$ g+g_{\leftmoon} -h + h_{\leftmoon}$ & $5.604\times10^{-8}$ \\
		$2 h - h_{\leftmoon}$ & $1.110\times10^{-6}$ &&$ g - g_{\leftmoon}+h - h_{\leftmoon}$  &$5.603\times10^{-8}$  \\
		$2g+2h -h_{\leftmoon} $ & $1.060\times10^{-6} $ &&$3 g - g_{\leftmoon}+2 h - h_{\leftmoon} $ & $5.449\times10^{-8}$ \\
		$2g-h+h_{\leftmoon}$ & $ 1.050\times10^{-6 }$ &&$g + g_{\leftmoon}-2 h + h_{\leftmoon} $& $5.155\times10^{-8}$ \\
		$ h_{\leftmoon}$& $5.311\times10^{-7} $& &$3 g + g_{\leftmoon}+h_{\leftmoon} $ &$3.975\times10^{-8}$\\
		$2g+ h+h_{\leftmoon}$& $4.018\times10^{-7}$ &&$3 g - g_{\leftmoon} -h_{\leftmoon} $ &$3.975\times10^{-8}$ \\
		$2g-2h$& $3.547\times10^{-7}$ &&$3g + g_{\leftmoon}- h + h_{\leftmoon} $& $2.262\times10^{-8}$\\

		$h+h_{\leftmoon} $ & $3.407\times10^{-7}$ &&$ 3 g + g_{\leftmoon}+ h + h_{\leftmoon} $& $2.162\times10^{-8}$ \\
		$2g-2h+h_{\leftmoon} $ & $1.546\times10^{-7}$ &&$ g -g_{\leftmoon}- h - h_{\leftmoon}$& $2.046\times10^{-8}$ \\
		$2g-h-h_{\leftmoon}$& $ 1.535\times10^{-7}$& &$g + g_{\leftmoon}+h+ h_{\leftmoon}$& $2.046\times10^{-8}$ \\
		$ 2h-2h_{\leftmoon}$ & $1.205\times10^{-7}$ & &$ g - g_{\leftmoon}+2 h - h_{\leftmoon}$&$1.971\times10^{-8}$ \\
		$2g+2h-2h_{\leftmoon}$&$ 1.150\times10^{-7} $ & &$g - g_{\leftmoon}+3 h - h_{\leftmoon} $& $1.755\times10^{-8}$  \\
	
		$2g+h-2h_{\leftmoon}$ &$5.899\times10^{-8}$ & &$3 g- g_{\leftmoon}+2 h - 2 h_{\leftmoon} $& $1.026\times10^{-8} $\\
		$h-2h_{\leftmoon}$&$5.004\times10^{-8}$ & &$ 3 g - g_{\leftmoon} +3 h - h_{\leftmoon} $& $1.001\times10^{-8}$\\ 
		$ 2h+h_{\leftmoon} $& $4.779\times10^{-8}$ & &&\\
			\noalign{\smallskip}\cline{1-2}	\cline{4-5}
\end{tabular}
\end{table}
Tab.~\ref{tab:ampsole} shows all the amplitudes of the second order solar harmonics. The third order solar harmonics computed are 28, but the corresponding coefficient are too small to be considered: the largest values are of the order of $10^{-11}\frac{km^2}{s^2}$ while the lowest ones are approximately $10^{-15}\frac{km^2}{s^2}$. \\ 
In the lunar case, the second order harmonics evaluated are 38, ranging from values of approximately $10^{-5}\frac{km^2}{s^2}$ to $10^{-11}\frac{km^2}{s^2}$. Instead, the third order contribution consists in 196 harmonics, ranging from approximately $10^{-8}\frac{km^2}{s^2}$ to $10^{-17} \frac{km^2}{s^2}$. Largest amplitudes of both the second and the third order lunar potential are listed in Tab.~\ref{tab:ampluna}.\\ 
Despite a Molniya satellite reaches high altitudes, the geocentric orbits of the Moon and of the Sun are nearly-circular and this fact may explain why both the lunar and the solar third order harmonics are quite small. In fact, $G_{31-1}(e)$ and $G_{321}(e)$, the eccentricity functions not vanishing for the third order expansions of the lunisolar doubly-averaged potential, are quite small for $e=e_{\leftmoon}, e_{\odot}$.  As already noticed in \cite{frontiers}, the third order contribution given by a third body with a circular orbit is null. \\
\begin{table}
	\caption{Largest amplitudes $\bigl[ \frac{rad}{s} \bigr]$, in absolute value, of the partial derivatives of the harmonic coefficients with respect to the actions, together with the corresponding  argument. The values are computed by evaluating the terms at the Molniya parameters.}
	\label{tab:deramp}
	\centering
	\begin{tabular}{lllll}
	\noalign{\smallskip}\cline{1-2} \cline{4-5}\noalign{\smallskip}
	$\varphi_{\alpha}^{\leftmoon}$, $l=2$& $|C_{\alpha}^{\leftmoon}\frac{\partial\mathcal{A}_{\alpha}^{\leftmoon}}{\partial G}(L_{moln},G_{moln},H_{moln})|$&   & $\varphi_{\alpha}^{\leftmoon}$, $l=2$& $|C_{\alpha}^{\leftmoon}\frac{\partial\mathcal{A}_{\alpha}^{\leftmoon}}{\partial H}(L_{moln},G_{moln},H_{moln})|$ \\
	\noalign{\smallskip}\cline{1-2} \cline{4-5}\noalign{\smallskip}
	Mean Term & $1.255\times 10^{-10}$   &    & Mean Term & $3.848\times10^{-10}$\\
	$2g+h$& $3.723\times10^{-10}$&   & $2g$ & $2.805\times 10^{-10}$ \\
	$2g$ & $3.406\times10^{-10}$ & & $h$ & $2.761\times10^{-10 }$ \\
	$h$ & $2.573\times10^{-10}$  &  & $2g-h$ & $1.757\times10^{-10}$\\                     
	
	$2g + 2h$& $8.436\times10^{-11}$ &    & $ h-h_{\leftmoon}$ & $ 5.478\times10^{-11}$ \\
	$2g+h - h_{\leftmoon}$& $7.384\times10^{-11}$ &   & $ h_{\leftmoon}$ & $ 4.994\times10^{-11}$ \\
	$2 g-h$& $5.924\times10^{-11}$ &    &$2g+2h$ & $4.708\times10^{-11}$ \\
	$h - h_{\leftmoon}$& $5.103\times10^{-11} $ &   &$2h$& $3.991\times10^{-11}$\\ 
	$2 g+2 h - h_{\leftmoon}$& $3.676\times10^{-11}$ &    & $2g-h+h_{\leftmoon} $ & $3.484\times10^{-11} $ \\
	$2g-h_{\leftmoon} $& $2.210\times10^{-11}$ &  & $2g+h$ & $2.560\times10^{-11} $ \\
	$2 g+h_{\leftmoon} $& $2.210\times10^{-11}$ &  & $ 2g+2h-h_{\leftmoon}$& $2.052\times10^{-11 } $ \\
	$h_{\leftmoon}$& $1.629\times10^{-11} $ &   &$2g+h_{\leftmoon}$ & $1.820\times10^{-11}$\\			
	$2 g-h + h_{\leftmoon} $& $1.175\times10^{-11}$  &     & $2g-h_{\leftmoon}$ & $1.820\times10^{-11}$\\
	$2h$& $1.115\times10^{-11}$ &   &$2h-h_{\leftmoon}$ & $1.739\times10^{-11}$\\
	$2g+h + h_{\leftmoon}$& $1.079\times10^{-11}$ &   &$2g-2h$  &  $1.798\times10^{-11}$ \\
	\noalign{\smallskip}\cline{1-2} \cline{4-5}\noalign{\smallskip}
	$\phi_{\gamma}^{\odot}$, $l=2$& $|C_{\gamma}^{\odot}\frac{\partial\mathcal{A}_{\gamma}^{\odot}}{\partial G}(L_{moln},G_{moln},H_{moln})|$&   & $\phi_{\gamma}^{\odot}$, $l=2$& $|C_{\gamma}^{\odot}\frac{\partial\mathcal{A}_{\gamma}^{\odot}}{\partial H}(L_{moln},G_{moln},H_{moln})|$ \\
	
	\noalign{\smallskip}\cline{1-2} \cline{4-5}\noalign{\smallskip}
	$2g+h-h_{\odot} $ & $1.724\times10^{-10}$&  & Mean Term & $1.781\times10^{-10} $ \\
	$2g$ & $ 1.577\times10^{-10}$&&  $2g$ & $ 1.299\times10^{-10}$ \\
	$h-h_{\odot}$& $1.191\times10^{-10} $ & & $h-h_{\odot}$& $1.278\times10^{-10}$  \\

	Mean Term& $5.810\times10^{-11}$&  & $2g-(h-h_{\odot})$ & $8.133\times10^{-11 }$ \\
	$2g + 2(h - h_{\odot})$ & $ 3.905\times10^{-11}$ & & $2g +2(h-h_{\odot})$ & $2.179\times 10^{-11}$\\
	$2g-(h - h_{\odot}) $& $2.742\times10^{-11}$&& $2(h-h_{\odot})$ & $1.848\times10^{-11 } $   \\
	& &  & $2g +h - h_{\odot}$ & $1.185\times10^{-11 }$ \\
	\noalign{\smallskip}\cline{1-2} \cline{4-5}
	\end{tabular}
\end{table}
In Tab.~\ref{tab:deramp} they are given the estimates of the largest values that the partial derivatives of the harmonic coefficients with respect to the actions can take in the Molniya region. This analysis would indicate the dominant terms in the angular dynamics defined by the last two equations in $\eqref{hamsys}$. Tab.~\ref{tab:deramp} on the left shows the lunisolar contribution to $\dot g$, while on the right they are reported the terms determining the dynamics of $\dot h$. Only the second order lunisolar contribution was taken into account. \\
By evaluating at the Molniya parameters the precession rate due to the oblateness
\begin{equation}\label{oblmoln}
\left\{\begin{array}{ll}
\vspace{0.13cm}
\dot{g}_{J_2}= \frac{\partial \mathcal{H}_{J_2}}{\partial G}(L_{moln},G_{moln},H_{moln})= 1.018\times10^{-11}\hspace{0.1cm}\frac{rad}{s}  \\
\dot{h}_{J_2}= \frac{\partial \mathcal{H}_{J_2}}{\partial H}(L_{moln},G_{moln},H_{moln})= -2.636\times10^{-8}\hspace{0.1cm}\frac{rad}{s}\\
\end{array}\right.
\end{equation}
 it is easy to note the consequences of an orbital inclination close to the critical inclination. The precession caused by the lunar mean term (Tab.~\ref{tab:deramp} on the left) is one order of magnitude larger than the oblateness one $\dot{g}_{J_2}$. The solar mean term produces a lower value with respect to the one of the Moon, but, it is still larger than the oblateness one. Moreover, also the amplitudes of the oscillations seem to be quite large. These facts imply that the third-body effects on the dynamics of the argument of perigee is small but not negligible if compared with the oblateness effect.\\
On the contrary, the partial derivatives of both the lunar and the solar mean terms on the right ensure that the oblateness effect is still the dominant perturbation affecting $\dot{h}$, as it usually happens in case of no frozen condition.\\
To better catch how the lunisolar perturbation may affect the angular dynamics, the \textit{periods} of the arguments involved in the doubly-averaged model $\eqref{hamiltonian}$ are computed by using $\dot{h}\approx \dot{h}_{J_2}$ and different approximations of $\dot{g}$ including all the lunisolar periodic terms with amplitude of oscillation $\geq10^{-10}$ and both the lunar and the solar mean terms as follows

 \begin{equation}\label{newapprox}
 \begin{aligned}
 \dot g(c_0,c_1) \approx &1.018\times10^{-11} + c_0\bigl[1.255\times 10^{-10}+ 5.810\times 10^{-11} \bigr]+\\
 & \hspace{1cm}+c_1\bigl[ -3.723\times 10^{-10}\cos(2g+h) 
+4.983\times10^{-10}\cos(2g) +2.573\times 10^{-10}\cos(h)\\
 &\hspace{1.7cm}+ 1.191\times10^{-10}\cos(h-h_{\odot})+1.724\times 10^{-10}\cos(2g+h-h_{\odot})\bigr]
 \end{aligned}
 \end{equation}

In Eq. $\eqref{newapprox}$, the parameters $c_i=0,1$ for $i=0,1$ are essentially used to switch the lunisolar major disturbance on or off, making a distinction between oscillating contribution and secular drifts. We can always fix the relative position between the ecliptic and the equatorial reference plane by choosing $h_{\odot}=0$. 
\begin{table}[h!]
	\caption{Values of $\dot{g}(c_0,c_1)$ $\big[\frac{rad}{s}\big]$, in the Molniya region, computed through $\eqref{newapprox}$ at the stable initial value of the argument of the perigee $g=270\hspace{0.1cm} \deg$ and at some values of the initial longitude of the ascending node $h$.}
	\label{tab:initialangles}
	\centering
	\begin{tabular}{llll}
		\hline\noalign{\smallskip}
		$c_0$ & $c_1$& \textbf{initial ascending node} & $\dot g(c_0,c_1)$ \\
		\noalign{\smallskip}\hline\noalign{\smallskip}
		0 &0& - & $1.018\times 10^{-11} $ \\
		1&0 & -&  $ 1.938\times10^{-10}$\\
		1 &1& $0 \deg$ & $6.622\times 10^{-10}$ \\
		1 &1& $45 \deg$ & $5.631 \times 10^{-10}$ \\
		1 &1& $90 \deg$ & $5.060\times 10^{-10}$ \\
		1 &1& $120 \deg$ & $5.097\times 10^{-10}$ \\
		1 &1&  $180 \deg$ & $6.079\times 10^{-10}$\\
		1 &1& $210 \deg$ & $6.761\times 10^{-10}$ \\
		1 &1&  $270\deg $ & $7.641\times 10^{-10}$\\
		1 &1& $340 \deg$ & $7.048\times 10^{-10}$ \\
		\noalign{\smallskip}\hline			
	\end{tabular}
\end{table}
To handle the periodic effects we need to use initial conditions also for the argument of pericenter and for the longitude of the ascending node of the satellite. An initial argument of perigee at $270 \hspace{0.1cm}\deg$ is the best stable condition due to the Russian latitude \cite{art}, thus, we focus on how lunisolar effects on the angles vary with respect to the initial ascending node of the satellite. This choice is dictated by the fact highlighted for instance by Anselmo and Pardini in \cite{anselmo}: the initial ascending node is crucial for the satellite lifetime. We adopt as different approximations of $\dot g$ significant values from the last column of Tab.~\ref{tab:initialangles}: 
\begin{equation}\label{gdotapprox}
\begin{aligned}
	&\dot{g}_{J_2}=+1.018\times 10^{-11}, \quad &\dot{g}_{0}= 1.938\times10^{-10} , \quad &\dot{g}_{1}=5.060\times 10^{-10}\\
	&\dot{g}_{2}=6.079\times 10^{-9}, \quad  &\dot{g}_{3}=6.622\times 10^{-10},	\quad & \dot{g}_{4}=7.641\times 10^{-10}
	\end{aligned}
\end{equation}
\begin{table}
	\caption{Largest periods ($> 7 \hspace{0.1cm}yr$) for the lunar and solar arguments appearing in the second order doubly-averaged disturbing potential. The first column helps to identify which third body the arguments belong to. The periods, measured in years $[yr]$, are computed with the approximations of $\dot g $ listed in Tab.~\ref{tab:initialangles} and $\dot h\approx\dot h_{J_2}$. The values $\dot g_{J_2}$, $\dot g_i$ for $i=0,1,2,3,4$ are detailed in Eq. $\eqref{gdotapprox}$.}
	\label{tab:periodi1}  
	\centering
	\begin{tabular}{llllllll}
		\hline\noalign{\smallskip}
&\textbf{Argument}& $\begin{array}{ll}
\text{\textbf{\textit{Period} with:}}\\
\left\{ \begin{array}{ll}
\dot g\approx \dot g_{J_2}\\
\dot h \approx \dot h_{J_2}
\end{array}\right.\end{array}$ &$\begin{array}{ll}
\text{\textbf{\textit{Period} with:}}\\
\left\{ \begin{array}{ll}
\dot g\approx \dot g_{0}\\
\dot h \approx \dot h_{J_2}
\end{array}\right.\end{array}$& $\begin{array}{ll}
\text{\textbf{\textit{Period} with:}}\\
\left\{ \begin{array}{ll}
\dot g\approx \dot g_{1}\\
\dot h \approx \dot h_{J_2}
\end{array}\right.\end{array}$ & $\begin{array}{ll}
\text{\textbf{\textit{Period} with:}}\\
\left\{ \begin{array}{ll}
\dot g\approx\dot g_{2}\\
\dot h \approx \dot h_{J_2}
\end{array}\right.\end{array}$& $\begin{array}{ll}
\text{\textbf{\textit{Period} with:}}\\
\left\{ \begin{array}{ll}
\dot g\approx\dot g_{3} \\
\dot h \approx \dot h_{J_2}
\end{array}\right.\end{array}$ & $\begin{array}{ll}
\text{\textbf{\textit{Period} with:}}\\
\left\{ \begin{array}{ll}
\dot g\approx\dot g_{4} \\
\dot h \approx \dot h_{J_2}
\end{array}\right.\end{array}$\\
\noalign{\smallskip}\hline\noalign{\smallskip}
	$\leftmoon,\odot$& $2g$ &9777.54 & 513.68 & 196.72& 163.76 & 150.31 & 130.27 \\
\noalign{\smallskip}\hline\noalign{\smallskip}
$\leftmoon$ & $2h_{\leftmoon} -h-2g$ & 40.25 & 43.47 & 50.34 & 53.07 & 54.65 & 57.89 \\
$\leftmoon$ & $2h_{\leftmoon}- h $ & 40.08 &  40.08  &  40.08  &  40.08 &  40.08 &  40.08  \\
$\leftmoon$ & $2h_{\leftmoon} -h+2g$ & 39.92 & 37.18 & 33.30 & 32.20 & 31.64 & 30.65 \\
\noalign{\smallskip}\hline\noalign{\smallskip}
$\leftmoon$ & $h_{\leftmoon} + 2g$ & 18.65 & 19.31 & 20.56 & 21.00 & 21.24 & 21.71 \\
$\leftmoon$ & $ h_{\leftmoon} $ & 18.61 & 18.61 & 18.61 & 18.61 & 18.61 & 18.61 \\ 
$\leftmoon$ & $h_{\leftmoon}-2g$ & 18.58 & 17.96 & 17.00 & 16.71 & 16.56 & 16.28 \\
\noalign{\smallskip}\hline\noalign{\smallskip}
$\leftmoon$ & $h_{\leftmoon} - h -2g $ &12.73  &13.03 & 13.58 & 13.78 & 13.88 & 14.08 \\
$\leftmoon$ & $ h_{\leftmoon} - h $ & 12.71 & 12.71 & 12.71 & 12.71 & 12.71 & 12.71 \\
$\leftmoon$ & $h_{\leftmoon} - h +2g$ &12.69 & 12.40 & 11.94 & 11.79  & 11.72 & 11.58 \\
\noalign{\smallskip}\hline\noalign{\smallskip}
$\leftmoon$ &$2h_{\leftmoon} + 2g$ & 9.31 &9.47 & 9.77 &  9.87 & 9.92 & 10.02\\
$\leftmoon$ & $2h_{\leftmoon}$ & 9.31 &  9.31 &   9.31 &   9.31 &   9.31 &   9.31 \\
$\leftmoon$ & $2h_{\leftmoon} - 2g$ & 9.30 & 9.14 & 8.89 & 8.81 & 8.76 & 8.68 \\
\noalign{\smallskip}\hline\noalign{\smallskip} 
$\leftmoon, \odot$ & $2g+h$ & 7.56 & 7.66 & 7.85 & 7.92& 7.95 & 8.01 \\
$\leftmoon, \odot$ & $h$ & 7.55 &7.55 & 7.55 & 7.55 & 7.55 & 7.55 \\
$\leftmoon, \odot$ & $2g-h$ & 7.55 & 7.55 & 7.27 & 7.21 & 7.19 & 7.13 \\
\noalign{\smallskip}\hline\noalign{\smallskip}  
	\end{tabular}
\end{table}
	\begin{table}
	\caption{Largest periods for the lunar and solar arguments appearing in the third order doubly-averaged disturbing potential. The first column helps to identify the third body. The periods, measured in years $[yr]$, are computed assuming the approximations of $\dot g $ listed in Tab.~\ref{tab:initialangles} and $\dot h=\dot h_{J_2}$. The values $\dot g_{J_2}$, $\dot g_i$ for $i=0,1,2,3,4$ are detailed in Eq. $\eqref{gdotapprox}$.}
	\label{tab:periodi2}
	\centering
		\begin{tabular}{llllllll}
					\hline\noalign{\smallskip}
				&\textbf{Argument}& $\begin{array}{ll}
				\text{\textbf{\textit{Period} with:}}\\
				\left\{ \begin{array}{ll}
				\dot g\approx \dot g_{J_2}\\
				\dot h \approx \dot h_{J_2}
				\end{array}\right.\end{array}$ &$\begin{array}{ll}
				\text{\textbf{\textit{Period} with:}}\\
				\left\{ \begin{array}{ll}
				\dot g\approx \dot g_{0}\\
				\dot h \approx \dot h_{J_2}
				\end{array}\right.\end{array}$& $\begin{array}{ll}
				\text{\textbf{\textit{Period} with:}}\\
				\left\{ \begin{array}{ll}
				\dot g\approx \dot g_{1}\\
				\dot h \approx \dot h_{J_2}
				\end{array}\right.\end{array}$ & $\begin{array}{ll}
				\text{\textbf{\textit{Period} with:}}\\
				\left\{ \begin{array}{ll}
				\dot g\approx\dot g_{2}\\
				\dot h \approx \dot h_{J_2}
				\end{array}\right.\end{array}$& $\begin{array}{ll}
				\text{\textbf{\textit{Period} with:}}\\
				\left\{ \begin{array}{ll}
				\dot g\approx\dot g_{3} \\
				\dot h \approx \dot h_{J_2}
				\end{array}\right.\end{array}$ & $\begin{array}{ll}
				\text{\textbf{\textit{Period} with:}}\\
				\left\{ \begin{array}{ll}
				\dot g\approx\dot g_{4} \\
				\dot h \approx \dot h_{J_2}
				\end{array}\right.\end{array}$\\
						\noalign{\smallskip}\hline\noalign{\smallskip}
				$\odot$ & $g-g_{\odot}$ & 23 669.36 & 1 036.82 & 394.83 & 328.48 & 301.43 & 261.16 \\
				$\odot$ & $g-g_{\odot}$ & 16 659.31  & 1 018.06 & 392.08 & 326.57 & 299.83 & 259.95 \\
				\noalign{\smallskip}\hline\noalign{\smallskip} 
				$\odot$ & $3g-g_{\odot}$ & 6 919.27 & 343.50 & 131.30 & 109.28 & 100.30 & 86.91 \\
				$\odot $ & $3g+g_{\odot}$ & 6 161.37 & 341.41 & 131.00 & 109.07 & 100.12 & 86.78 \\
				\noalign{\smallskip}\hline\noalign{\smallskip}
				$\leftmoon$ & $3g-g_{\leftmoon} -3h_{\leftmoon}$ & 184.42 & 367.55 & 488.04& 279.03 & 227.11 & 168.40 \\
				$\leftmoon$ & $g-g_{\leftmoon}-3h_{\leftmoon}$ & 181.00 & 217.27 & 329.57 & 396.41 & 444.55 & 575.42 \\
				$\leftmoon$ & $g+g_{\leftmoon} +3h_{\leftmoon}$ & 177.71 & 152.69 & 123.19 & 115.89 & 112.33 & 106.23 \\
				$\leftmoon$ & $3g + g_{\leftmoon}+3h_{\leftmoon}$ & 174.54 & 117.70 & 75.75 & 67.86 & 64.29 & 58.51 \\
				\noalign{\smallskip}\hline\noalign{\smallskip}
				$\leftmoon$ & $3g -g_{\leftmoon} -2h +2h_{\leftmoon}$ & 108.11 & 154.24 & 562.28 &4 104.67 & 1 736.91 & 473.80 \\
				$\leftmoon$ & $ g -g_{\leftmoon} -2h +2h_{\leftmoon}$ & 106.93 & 118.62 & 145.73 & 157.48 & 164.55 & 179.68 \\
				$\leftmoon$ & $g +g_{\leftmoon} +2h -2h_{\leftmoon}$ & 105.77 & 96.37 & 83.72 & 80.28 & 78.55 & 75.52 \\
				$\leftmoon$ & $3g +g_{\leftmoon} +2h -2h_{\leftmoon}$ & 104.64 & 81.15 & 58.72 & 53.87 & 51.59 & 47.80 \\
				\noalign{\smallskip}\hline\noalign{\smallskip}
				$\leftmoon$ & $3g + g_{\leftmoon}+h +h_{\leftmoon} $ & 52.03 & 60.78 & 85.12 & 97.91 & 106.45 & 127.23 \\
				$\leftmoon$ & $g + g_{\leftmoon}+h +h_{\leftmoon} $ & 51.75 & 54.35 & 59.41 & 61.27 & 62.32 & 64.37 \\
				$\leftmoon$ & $g - g_{\leftmoon}-h -h_{\leftmoon} $ & 51.48 & 49.15 & 45.63 & 44.59 & 44.05 & 43.08 \\
				$\leftmoon$ & $3g - g_{\leftmoon}-h -h_{\leftmoon} $ & 51.21 & 44.86 & 34.07 & 37.03 & 35.04 & 32.37 	\\
				\noalign{\smallskip}\hline\noalign{\smallskip}
		\end{tabular}
\end{table}
In Tab.~\ref{tab:periodi1} they are collected the largest second order periods obtained, while Tab.~\ref{tab:periodi2} shows the third order ones. In both tables the arguments are grouped with respect to the associated periods to make the reading easier. In Tab.~\ref{tab:periodi1}, macro periods of the order $40.08\hspace{0.1cm}yr$, $18.61\hspace{0.1cm}yr$, $12.71\hspace{0.1cm}yr$, $9.30\hspace{0.1cm}yr$ and $7.55\hspace{0.1cm}yr$ are highlighted. In particular, we find the well-known value $18.61\hspace{0.1cm}yr$ in correspondence with the period of the lunar ascending node. Small periods are related to high frequencies which are not very sensitive to the value of $\dot g$; on the contrary, the largest periods strongly depend on the approximation chosen. The same feature also emerges from Tab.~\ref{tab:periodi2} where the groups are of four arguments in the lunar case and of two arguments in the solar case.\\
The argument $2g$ represents the \textit{main resonant angle}, because of the critical inclination: the oblateness approximation (Tab.~\ref{tab:periodi1}, first column) leads to a clearly huge period, indeed. By \textit{increasing} the lunisolar perturbation, the period decreases, although it still remains quite large. \\
The solar third order critical arguments $g\pm g_{\odot}$ and $3g\pm g_{\odot}$ (top of Tab.~\ref{tab:periodi2}) behave as the main resonant angle. Conversely, the third order lunar arguments $3g-g_{\leftmoon}-2h +2h_{\leftmoon}$ and $3g+g_{\leftmoon} +h+h_{\leftmoon}$ behave in the opposite way: by increasing the lunisolar perturbation the arguments may become even critical.\\
\begin{table}[h!]
	\caption{Lunisolar harmonics with \textit{ratio}$> 1$. The first column helps to identify the third body, the second one indicates if the corresponding argument appears in the second or in the third order lunisolar potential. The \textit{ratio}, measured in $\frac{km^2}{s}$, are computed by assuming the approximations of $\dot g$ listed in Tab.~\ref{tab:initialangles} and $\dot h=\dot h_{J_2}$. The values $\dot g_{J_2}$, $\dot g_i$ for $i=0,1,2,3,4$ are detailed in Eq. $\eqref{gdotapprox}$.}
	\label{tab:ratio}
	\centering
	\begin{tabular}{lllllllll}
		\hline\noalign{\smallskip}
		&$l$&\textbf{Argument} &  $\begin{array}{ll}
		\text{\textbf{\textit{Ratio}:}}\\
		\left\{ \begin{array}{ll}
		\dot g\approx \dot g_{J_2}\\
		\dot h \approx \dot h_{J_2}
		\vspace{0.12cm}
		\end{array}\right. \end{array}$&   $\begin{array}{ll}
		\text{\textbf{\textit{Ratio}:}}\\
		\left\{ \begin{array}{ll}
		\dot g\approx \dot g_{0}\\
		\dot h \approx \dot h_{0}
		\vspace{0.12cm}
		\end{array}\right. \end{array}$  & $\begin{array}{ll}
		\text{\textbf{\textit{Ratio}:}}\\
		\left\{ \begin{array}{ll}
		\dot g\approx\dot g_{1} \\
		\dot h \approx \dot{h}_{1}
		\vspace{0.12cm}
		\end{array}\right.\end{array}$ &  $\begin{array}{ll}
		\text{\textbf{\textit{Ratio}:}}\\
		\left\{ \begin{array}{ll}
		\dot g\approx\dot g_{2} \\
		\dot h \approx \dot{h}_{2}
		\vspace{0.12cm}
		\end{array}\right.\end{array}$&  $\begin{array}{ll}
		\text{\textbf{\textit{Ratio}:}}\\
		\left\{ \begin{array}{ll}
		\dot g\approx\dot g_{3} \\
		\dot h \approx \dot{h}_{3}
		\vspace{0.12cm}
		\end{array}\right.\end{array}$ &  $\begin{array}{ll}
		\text{\textbf{\textit{Ratio}:}}\\
		\left\{ \begin{array}{ll}
		\dot g\approx\dot g_{4} \\
		\dot h \approx \dot{h}_{4}
		\vspace{0.12cm}
		\end{array}\right.\end{array}$\\
		\hline\noalign{\smallskip}
		$\leftmoon$& 2&$ 2g $&  879 496.40  & 46 205.55 & 17 695.51 &14 730.36 &13 520. 88 & 11 718.5 \\
		$\odot$ &2&$2 g$& 407 137.87 & 21 389.55 &  8 191.64& 6 819.00 & 6 259.11 & 5 424.75   \\
		$\leftmoon$&2&$2 g + h$& 526.48 & 533.92& 547.08 &551.51& 553.90 & 558.45 \\
		$\leftmoon$&2&$h$& 446.00 & 446.00 & 446.004  & 446.00 & 446.00 & 446.00\\
		$\odot$&2&$2g+h$& 243.72 & 247.16  & 253.25& 255.30& 256.41 & 258.52  \\
		$\odot$&2	& $h$& 206.46 & 206.46  & 206.46& 206.46&206.46 & 206.46 \\
		$\leftmoon$&2&$2 g - h$& 200.75 & 197.99 &  193.47&192.05 & 191.29 & 189.89 \\
		$\leftmoon$&2&$2g+h - h_{\leftmoon} $& 175.79 &180.01  & 187.69& 190.33& 191.78 & 194.54  \\
		$\leftmoon$&2&$h-h_{\leftmoon} $& 148.84 & 148.84  &  148.84 & 148.84 & 148.84 &148.84   \\
		$\leftmoon$&2&$2g+h_{\leftmoon} $& 108.85 & 112.73    & 120.00& 122.58 & 124.00 & 126.75 \\		
		$\leftmoon$&2&$2g-h_{\leftmoon} $& 108.44& 104.85   &99.25& 97.56&96.67 & 95.06  \\
		$\odot$&2 &$  2 g - h $& 92.93 & 91.65  & 89.56& 88.90 & 88.55 & 87.90  \\
		$\leftmoon$&2&$2 g - h+h_{\leftmoon} $&66.96 & 65.43  & 62.97& 62.22&  61.82 & 61.08 \\
					 
		$\leftmoon$&2	&$ h_{\leftmoon}$& 49.65 &  49.65 & 49.65 & 49.65&49.65 & 49.65 \\
		$\leftmoon$&2	&$2 h$& 48.33 & 48.33   & 48.33& 48.33 & 48.33 &48.33   \\
		$\leftmoon$&2	&$2 (g + h)$& 46.15 & 46.48  &47.04& 47.22& 47.32 & 47.51   \\
		$\leftmoon$&2	&$2 h -h_{\leftmoon}$&26.42&  26.42   & 26.42& 26.42& 26.42 & 26.42 \\
		$\leftmoon$&2	&$2 (g + h)-h_{\leftmoon}$& 25.23 & 25.46&25.84& 25.97 & 26.04& 26.17  \\
		$\odot$&2&$2 h$& 22.37  &22.3 & 22.37&22.37 & 22.37 & 22.37 \\
		$\odot$&2&$2 ( g + h)$& 21.36&  21.51  & 21.77& 21.86&21.91 & 21.99  \\			
		$\leftmoon$&2&$2 h_{\leftmoon} - 2 g - h$& 11.92 & 12.88 & 14.91 & 15.72 & 16.19 & 17.15 \\
		$\leftmoon$&2&$h_{\leftmoon} + 2 g + h$& 10.85 & 10.96  & 11.15 & 11.21& 11.24 & 11.31 \\
		$\leftmoon$&2&$2 h_{\leftmoon} - h$& 10.07& 10.07 & 10.07& 10.07&10.07 & 10.07 \\
		$\leftmoon$&2&$h_{\leftmoon} + h$& 9.19 & 9.19   & 9.19& 9.19 & 9.19  & 9.19  \\
		$\leftmoon$&2&$2g - 2h$& 6.72&  6.68  & 6.60 & 6.58 & 6.56 & 6.54  \\
		$\leftmoon$&3 &$3 g + g_{\leftmoon} + h + h_{\leftmoon}$& 5.65&6.60& 9.24& 10.63& 11.56& 13.82 \\
		$\leftmoon$&3&$g + g_{\leftmoon} + h + h_{\leftmoon}$& 5.32& 5.58& 6.10 &6.29& 6.40 & 6.61\\
		$\leftmoon$&3&$g - g_{\leftmoon} - h - h_{\leftmoon}$ &5.29& 5.05&4.69 &4.58&4.53 & 4.43 \\
		$\leftmoon$&2&$2 h_{\leftmoon} + 2 g - h$&4.51&4.21 & 3.77& 3.64 & 3.58 & 3.47  \\		
		$\leftmoon$&2&$h_{\leftmoon} - 2 g + h$& 4.14& 4.10& 4.03 & 4.01 & 4.00 & 3.98   \\			
		$\leftmoon$&2&$2 h_{\leftmoon} -2 h$& 3.85 & 3.85  & 3.85& 3.85 &3.85  & 3.85 \\
		$\leftmoon$&2&$h_{\leftmoon} + 2 g - 2 h$&3.68 & 3.64 & 3.59& 3.57 & 3.57 & 3.56   \\
		$\leftmoon$&2&$2 (h_{\leftmoon} - g - h)$& 3.67& 3.72& 3.79& 3.82& 3.83 & 3.86  \\
		$\odot$&2 & $2 ( g - h)$& 3.11 &  3.09& 3.06&3.04& 3.04 & 3.03  \\
		$\leftmoon$&3&$3 g - g_{\leftmoon} - h - h_{\leftmoon}$ & 2.12& 1.86 & 1.53& 1.45& 1.41 & 1.34\\
		$\leftmoon$&3&$3 g - g_{\leftmoon} - h_{\leftmoon}$& 1.77&1.81&1.89& 1.92& 1.94 & 1.97  \\
		$\leftmoon$&3&$3 g + g_{\leftmoon} + h_{\leftmoon}$& 1.76& 1.18&1.65&1.63& 1.62 & 1.60\\
		$\leftmoon$&3&$3 g + g_{\leftmoon} + h $ & 1.27&1.18&1.05&1.01& 0.99& 0.96\\
		$\leftmoon$&3& $g - g_{\leftmoon} - h$& 1.21& 1.25& 1.31&1.33&1.34 & 1.36\\
		$\leftmoon$&3&$3 g - g_{\leftmoon} + h - h_{\leftmoon}$& 1.21& 1.23&1.25& 1.26& 1.26 & 1.27\\
		$\leftmoon$&3& $g + g_{\leftmoon} + h$ & 1.21&1.18&1.13&1.11&1.10 & 1.09 \\
	   $\leftmoon$&3&$g - g_{\leftmoon} + h - h_{\leftmoon}$& 1.15 & 1.15&1.16&1.16& 1.16 & 1.16 \\
		$\leftmoon$&3&$g + g_{\leftmoon}- h + h_{\leftmoon}$& 1.15&1.14&1.13 &1.13& 1.13 & 1.13\\
		\hline\noalign{\smallskip}
	\end{tabular}
\end{table}
In Tab.~\ref{tab:ratio} all the harmonics whose \textit{ratio} is greater than $1$ are shown, from the highest to the lowest with respect to the fourth column. On the top of the table we find the main resonant argument $2g$. The amplitude of this term is quite large both as lunar (see Tab.~\ref{tab:ampluna}) and as solar harmonics (see Tab.~\ref{tab:ampsole}); moreover, the period is huge because of the critical inclination. Clearly, $2g$ is the lunisolar term dominating the Molniya dynamics: it is known that its periodic component originates the deepest growth in eccentricity on a long-term timescale.\\
Subsequently, the harmonics corresponding to $2g\pm h$ and $h$ seem to produce a significant contribution to the dynamics. These arguments are far from being critical, the periods are around $7.55 \hspace{0.1cm}yr$, but their amplitudes are quite large (Tabs.~\ref{tab:ampluna} and ~\ref{tab:ampsole}). In \cite{cinesimoln} a pure numerical orbit, computed by using the observational data from the Two-Line Element (TLE), was compared with the results given by a double resonances model with $2\dot g\pm \dot h$ and $2g$ which qualitatively catches the main characteristic of the long-term evolution of $\omega$, $e$ and $i$.\\
Also, it has to be noted that three third order lunar harmonics show a \textit{ratio} larger than few second order terms; their amplitudes are of the order of $10^{-8} \frac{km^2}{s^2}$, according to Tab.~\ref{tab:ampluna}. In particular, the increase of the \textit{ratio} corresponding to $3g+g_{\leftmoon}+h + h_{\leftmoon}$ in the last four columns is directly related with the growth of its period shown in Tab.~\ref{tab:periodi2}. \\
Finally, from Tab.~\ref{tab:ratio} we can conclude that the second order lunisolar effect is the dominant perturbation on the long-term dynamics, as already found numerically in \cite{molnarxiv}. \\

\subsection{The phase space structure of resonances}
\label{sec:phasespaceres}
\begin{table}
	\caption{Resonances whose dynamics is well described by the SRM. The first column identifies the resonances through the critical argument associated with, the second column shows the first integral arising from the resonant Hamiltonian. $\Gamma$ is the dummy momentum introduced to add one dimension to the phase space in case of resonances involving the lunar ascending node. The center of libration is given in terms of Keplerian elements: $(e^*, i^*)$ identifies the libration center of the exact resonance while $(e_s,i_s)$ gives the center of libration related to the real resonance. $|J_{1s}- J_{1u}|$ gives information about the asymmetry of the resonant region (see Sect.~\ref{sec:1resdynamics}). The last two columns show the libration width in terms of $e$ and $i$, the same values are obtained with the SRM Eq. $\eqref{maxlibwidth}$ and with NSA Eq. $\eqref{width}$.}
	\label{tab:SRMok} 
	\centering
	\begin{tabular}{lllllllll}
		\hline\noalign{\smallskip}
		\textbf{Critical Argument} & \textbf{First Integral}& $e^*$& $e_s$ & $i^*$ $\deg$ &$i_s$ $\deg$& $|J_{1s}- J_{1u}|$ $\frac{km^2}{s}$ & $\Delta e$ & $\Delta i$ $\deg$ \\
		\noalign{\smallskip}\hline\noalign{\smallskip}	\noalign{\smallskip}
	$2g-h $ & $\sqrt{\mathnormal{a}(1-e^2)}(\cos i + \frac{1}{2})$ &0.64 & 0.64& 69.14 & 69.03 & 114.41&0.13 & 7.3\\
	\noalign{\smallskip}	\noalign{\smallskip}
	$2g+h$ & $\sqrt{\mathnormal{a}(1-e^2)}(\frac{1}{2}-\cos i)$ &0.98 &0.98&56.06&56.06&1.20  &0.004&0.55\\
	\noalign{\smallskip}	\noalign{\smallskip}
	$2g+h-h_{\leftmoon}$ & $\begin{aligned} &\Gamma - \frac{1}{2}\sqrt{\mathnormal{a}(1-e^2)} \\
	&\sqrt{\mathnormal{a}(1-e^2)}\cos i\end{aligned} $ &0.52 &0.52& 62.79& 62.80&337.35 &0.15&0.29\\ 
	\noalign{\smallskip}	\noalign{\smallskip}
	$2g + h_{\leftmoon}$ & $\begin{aligned} &\Gamma + \frac{1}{2}\sqrt{\mathnormal{a}(1-e^2)} \\
	&\sqrt{\mathnormal{a}(1-e^2)}\cos i\end{aligned}$ &0.76 &0.76 & 61.56&61.55&10.27 &0.03&1.63 \\ 
		\noalign{\smallskip}\hline
	\end{tabular}
\end{table}
\begin{figure*}
	\centering
	\begin{minipage}{0.35\textwidth}
	\includegraphics[width=0.75\textwidth]{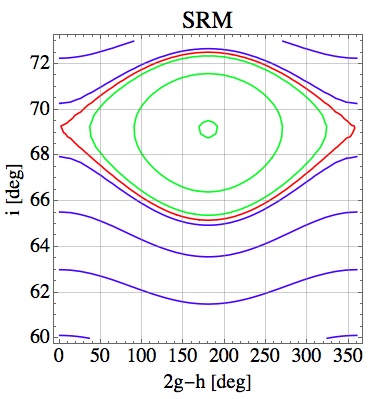}\\
\end{minipage}%
\begin{minipage}{0.35\textwidth}
	\includegraphics[width=0.75\textwidth]{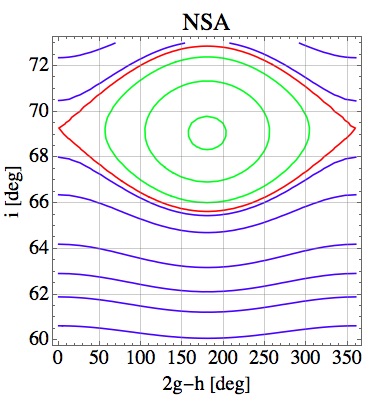}\\
\end{minipage}\\
	\begin{minipage}{0.35\textwidth}
	\includegraphics[width=0.75\textwidth]{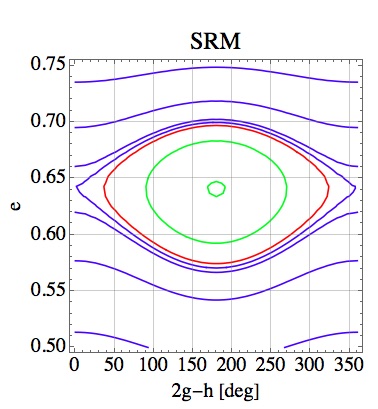}\\
\end{minipage}%
\begin{minipage}{0.35\textwidth}
	\includegraphics[width=0.75\textwidth]{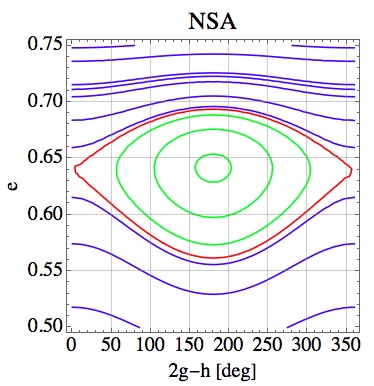}\\
\end{minipage}
	\caption{$2\dot g-\dot h$ resonance: contour plot of the pendulum-like approximation (SRM on the left) and of the resonant Hamiltonian not developed in Taylor series (NSA on the right). The X-axis always shows the critical angle $2g-h $ in $\deg$. The Y-axis is converted in $i$ (on the top), and measured in $\deg$, or in $e$ (on the bottom). \textit{Green} lines denote the librating curves around the stable equilibrium, \textit{blue} lines denote the circulation region while the separatrices are sketched in \textit{red}. }
	\label{fig:2gmh}       
\end{figure*}
\begin{figure*}
	\centering
	\begin{minipage}{0.35\textwidth}
	\includegraphics[width=0.75\textwidth]{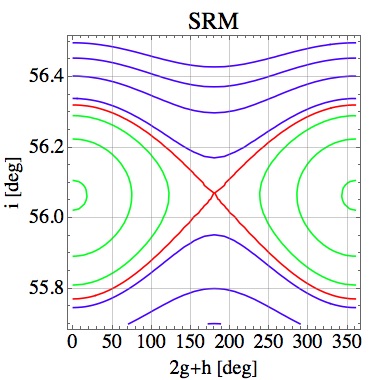}\\
\end{minipage}%
\begin{minipage}{0.35\textwidth}
	\includegraphics[width=0.75\textwidth]{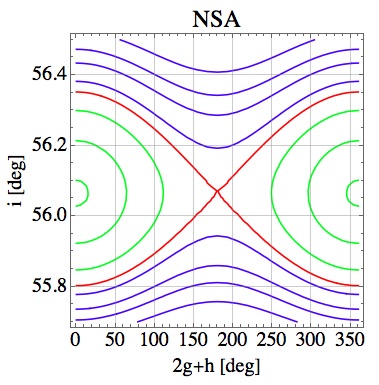}\\
\end{minipage}\\
\begin{minipage}{0.35\textwidth}
	\includegraphics[width=0.75\textwidth]{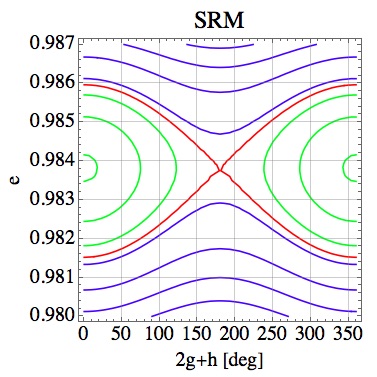}\\
\end{minipage}%
\begin{minipage}{0.35\textwidth}
	\includegraphics[width=0.75\textwidth]{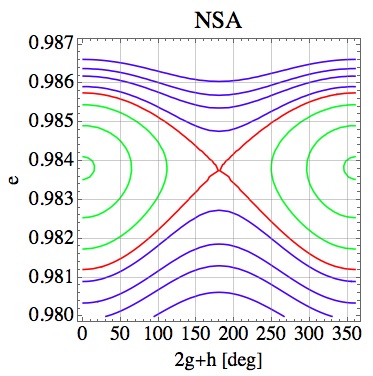}\\
\end{minipage}
	\caption{$2\dot g+\dot h$ resonance: contour plot of the pendulum-like approximation (SRM on the left) and of the resonant Hamiltonian not developed in Taylor series (NSA on the right). The X-axis always shows the critical angle $2g+h$ in $\deg$. The Y-axis is converted in $i$ (on the top), and measured in $\deg$, or in $e$ (on the bottom).  \textit{Green} lines denote the librating curves around the stable equilibrium, \textit{blue} lines denote the circulation region while the separatrices are sketched in \textit{red}.}
	\label{fig:2gph}       
\end{figure*}
\begin{figure*}
	\centering
	\begin{minipage}{0.35\textwidth}
		\includegraphics[width=0.75\textwidth]{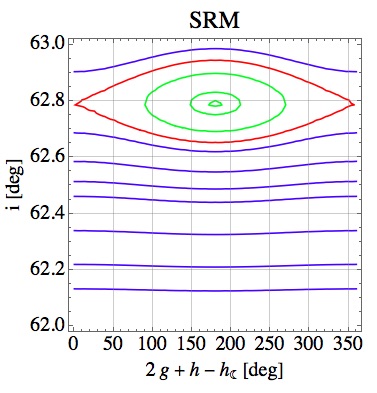}\\
	\end{minipage}%
	\begin{minipage}{0.35\textwidth}
		\includegraphics[width=0.75\textwidth]{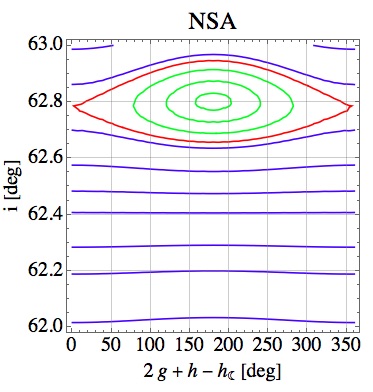}\\
	\end{minipage}\\
	\begin{minipage}{0.35\textwidth}
		\includegraphics[width=0.75\textwidth]{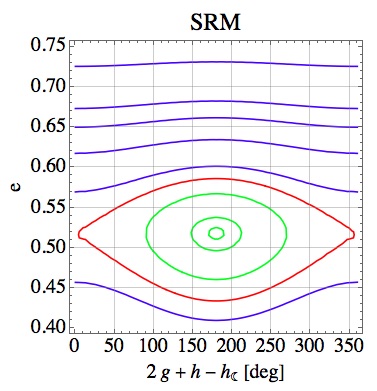}\\
	\end{minipage}%
	\begin{minipage}{0.35\textwidth}
		\includegraphics[width=0.75\textwidth]{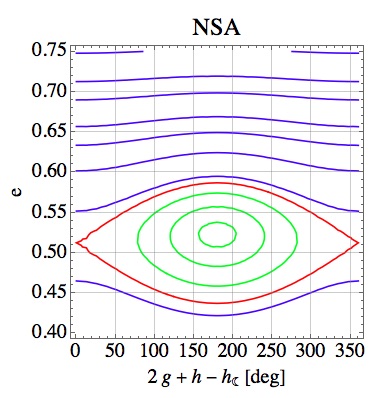}\\
	\end{minipage}
	\caption{$2\dot g+\dot h-\dot{h}_{\leftmoon}$resonance: contour plot of the pendulum-like approximation (SRM on the left) and of the resonant Hamiltonian not developed in Taylor series (NSA on the right). The X-axis always shows the critical angle $2g+h-h_{\leftmoon}$ in $\deg$. The Y-axis is converted in $i$ (on the top), and measured in $\deg$, or in $e$ (on the bottom). \textit{Green} lines denote the librating curves around the stable equilibrium, \textit{blue} lines denote the circulation region while the separatrices are sketched in \textit{red}.}
	\label{fig:2gphmhm}       
\end{figure*}
\begin{figure*}
	\centering
	\begin{minipage}{0.35\textwidth}
		\includegraphics[width=0.8\textwidth]{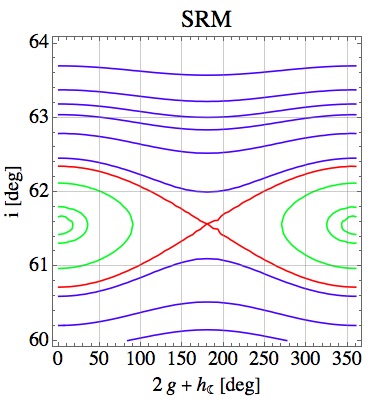}\\
	\end{minipage}%
	\begin{minipage}{0.35\textwidth}
		\includegraphics[width=0.8\textwidth]{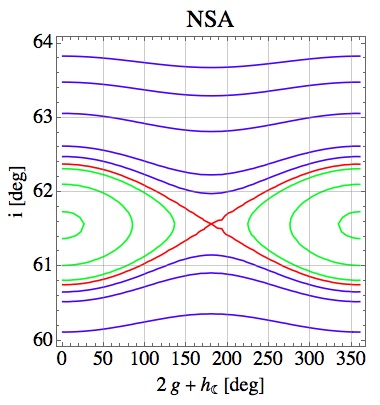}\\
	\end{minipage}\\
	\begin{minipage}{0.35\textwidth}
		\includegraphics[width=0.8\textwidth]{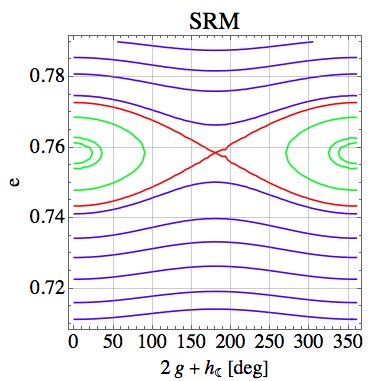}\\
	\end{minipage}%
	\begin{minipage}{0.35\textwidth}
		\includegraphics[width=0.8\textwidth]{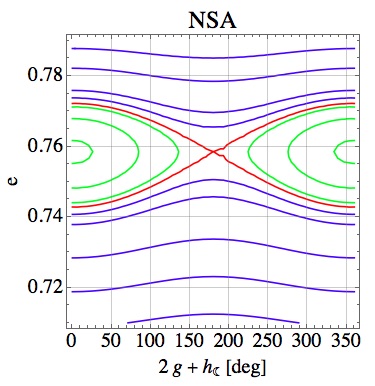}\\
	\end{minipage}
	\caption{$2\dot g+\dot{h}_{\leftmoon}$ resonance: contour plot of the pendulum-like approximation (plot title SRM on the left) and of the resonant Hamiltonian not developed in Taylor series (plot title NSA on the right). The X-axis always shows the critical angle $2g+h_{\leftmoon}$ in $\deg$. The Y-axis is converted in inclination (on the top), and measured in $\deg$, or in eccentricity (on the bottom).  \textit{Green} lines denote the librating curves around the stable equilibrium, \textit{blue} lines denote the circulation region while the separatrices are sketched in \textit{red}.}
	\label{fig:2gphm}   
\end{figure*}
\begin{figure*}
	\centering
	\begin{minipage}{0.35\textwidth}
		\includegraphics[width=0.8\textwidth]{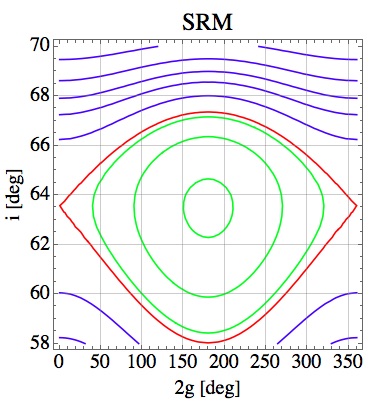}\\
	\end{minipage}%
	\begin{minipage}{0.35\textwidth}
		\includegraphics[width=0.8\textwidth]{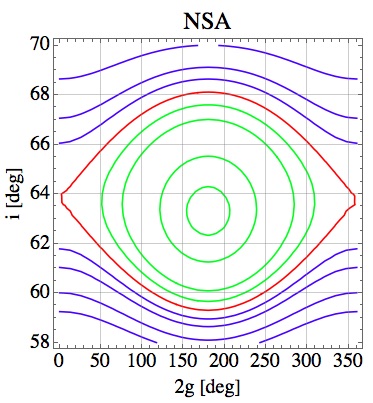}\\
	\end{minipage}\\
	\begin{minipage}{0.35\textwidth}
		\includegraphics[width=0.8\textwidth]{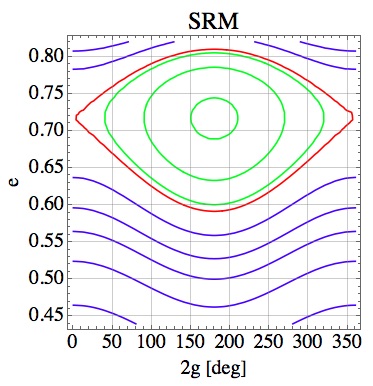}\\
	\end{minipage}%
	\begin{minipage}{0.35\textwidth}
		\includegraphics[width=0.8\textwidth]{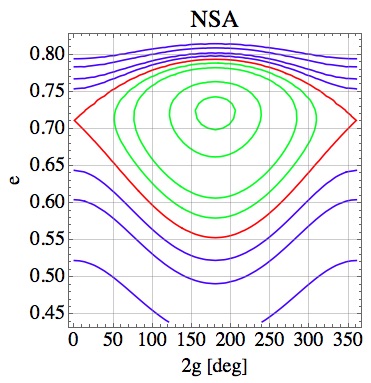}\\
	\end{minipage}
	\caption{$2\dot g$ resonance: contour plot of the pendulum-like approximation (plot title SRM on the left) and of the resonant Hamiltonian not developed in Taylor series (plot title NSA on the right). The X-axis always shows the critical angle $2g$ in $\deg$. The Y-axis is converted in inclination (on the top), and measured in $\deg$, or in eccentricity (on the bottom).  \textit{Green} lines denote the librating curves around the stable equilibrium, \textit{blue} lines denote the circulation region while the separatrices are sketched in \textit{red}.}
	\label{fig:2g}       
\end{figure*}
We are interested in the dynamical behaviour around the lunisolar dominant harmonics, thus we follow the theoretical discussion in Sect.~\ref{sec:1resdynamics} with: $\mathbf{I}=(L,G,H)$, $\bm{\psi}=(\ell,g,h)$, the unperturbed term $\mathcal{H}_0$ is given by the $J_2$-term and both the lunar and solar mean terms, that is: 
\begin{equation}\label{unpres}
\mathcal{H}_{0}= \mathcal{H}_{J_2}(G,H;L) + C_{0}^{\leftmoon}\mathcal{A}_{0}^{\leftmoon}(G,H;L) +  C_{0}^{\odot}\mathcal{A}_{0}^{\odot}(G,H;L) 
\end{equation}
The resonant perturbation is given by the Hamiltonian contribution of a lunisolar dominant harmonic.
Under the hypothesis of isolated resonance, the dynamics in a small enough neighborhood of a particular resonance is described by a resonant Hamiltonian of the form given in Eq. $\eqref{resham}$. 
$L$ is a first integral and we focus on the level curve $L=L_{moln}$. 
The resonant dominant harmonics for which the SRM gives a reliable description of the phase plane structure are listed in Tab.~\ref{tab:SRMok}, while, the resonances associated with the arguments in Tab.~\ref{tab:SRMno} exhibit a non-standard behaviour. \\
According to Sect.~\ref{sec:1resdynamics}, after performing a coordinate change of the form shown in Eq. $\eqref{resvar}$, the motion evolves in the $(J_1,\theta_1)$ plane. Except for the polar resonance $\dot h$, the first integral is Kozai-like, that is, the evolution of $e$ and $i$ is coupled because the semi-major axis is constant in the long-term \cite{morbidelli}. \\
Since the harmonic argument depending on the lunar ascending node leads to a non-autonomous resonant Hamiltonian, it is necessary to introduce a dummy momentum $\Gamma$ and a new conjugate angle depending on the lunar node in order to eliminate the explicit linear time dependency, e.g. \cite{meoreg}. For this reason, there are two first integrals in correspondence of $2g\pm h_{\leftmoon}$, $2g+h-h_{\leftmoon}$.\\
A first integral constrains the motion, thus all the results depend on the initial conditions used to evaluate the conserved quantity; if not specified, we assume the Molniya parameters.
In Tab.~\ref{tab:SRMok} it is pointed out the center of libration related to each resonant harmonic and the corresponding maximum width, as computed with the standard approach (SRM) through Eq. $\eqref{maxlibwidth}$. The maximum real excursion in eccentricity and in inclination, computed by using Eq. $\eqref{width}$, gives substantially the same width obtained with the SRM. These facts can be appreciated by looking at the phase portraits from Figs.~\ref{fig:2gmh}-~\ref{fig:2gphm}. The dynamical structure arising from the pendulum-like Hamiltonian is depicted on the left, while on the right they are shown the results obtained from the resonant Hamiltonian not developed in Taylor series; the Y-axis is always converted in eccentricity or in inclination. 
\paragraph{The resonance $2\dot g+\dot h$}
By using the Molniya parameter as initial conditions, the feasible equilibrium lies in the retrograde orbit region, at $i=110.99\hspace{0.1cm}\deg$. To obtain the resonant region of $2\dot g+\dot h$ (see Fig.~\ref{fig:2gph}) around the well-known inclination of approximately $56\hspace{0.1cm}\deg$, that is to find the equilibria of the corresponding system in the prograde orbit environment, it was necessary to consider a different initial condition: instead of $i=i_{moln}$ as initial inclination, we have adopted $i=59\hspace{0.1cm}\deg$. 
It means that for a Molniya satellite with $(\mathnormal{a}_{moln},e_{moln},i_{moln})$ as initial condition the argument $2g+h$ always circulate with a period of approximately $7.6\hspace{0.1cm}yr$ (see Tab.~\ref{tab:periodi1}). In any case, the libration region of $2\dot g+\dot h$ is quite narrow and do not overlap with the other resonances taken into account, especially with the main resonance and with $2\dot g -\dot h $ as already found in \cite{cinesimoln}. 
\begin{table}
	\caption{Resonances whose dynamics is not appropriately described by the SRM. The first column identifies the resonance through the associated argument, the second column shows the first integral arising from the resonant Hamiltonian. The equilibria are given in terms of eccentricity and inclination: $(e_s,i_s)$ stable ones, $(e_u,i_u)$ unstable ones. $|J_{1s}- J_{1u}|$ gives information about the asymmetry of the resonant region (see Sect.~\ref{sec:1resdynamics}). The last two columns show the maximum excursion, in terms of $e$ and $i$, that may be attained in the libration region as computed with NSA through Eq. $\eqref{width}$. For the last resonance reported, the values correspond to a bifurcation, see the text for more details.}
	\label{tab:SRMno} 
	\centering
	\begin{tabular}{lllllllll}
		\hline\noalign{\smallskip}
		\textbf{Critical Argument} & \textbf{First Integral}& $e_s$& $e_u$ & $i_s$ $\deg$ &$i_u$ $\deg$& $|J_{1s}- J_{1u}|$ $\frac{km^2}{s}$ & $[e_{min},e_{max}]$ & $[i_{min},i_{max}]$ $\deg$ \\
		\noalign{\smallskip}\hline\noalign{\smallskip} \noalign{\smallskip}
		$2g$ & $\sqrt{\mathnormal{a}(1-e^2)}\cos i$ & 0.72  &0.71  &63.29 & 63.69 & 625.10 & $[0.55,0.79]$ & $[59.30, 68.11]$ \\
		\noalign{\smallskip}\noalign{\smallskip}
		$h$ & $e$& 0.72 & 0.72 & $\begin{aligned} &89.43\\ &90.57 \end{aligned}$  & $\begin{aligned}
		&90 \\
		&90 
		\end{aligned}$ & not evaluated & - &$\begin{aligned}
		&[88.85,90.00]\\
		&[90.00,91.14]\end{aligned}$\\ 
		\noalign{\smallskip}\noalign{\smallskip}
		$2g-h_{\leftmoon}$ &$\begin{aligned} &\sqrt{\mathnormal{a}(1-e^2)}\cos i\\ &\Gamma+\frac{1}{2}\sqrt{\mathnormal{a}(1-e^2)} \end{aligned}$&$\begin{aligned} &0.62\\ &0.57 \end{aligned}$   & $\begin{aligned} &0.60\\ &0.54 \end{aligned}$ & $\begin{aligned} &67.97\\ &68.95 \end{aligned}$ & $\begin{aligned} &68.48\\ &69.57 \end{aligned}$ & not evaluated &$[0.44,0.68]$ & $[66.28,70.83]$ \\
		\noalign{\smallskip}\noalign{\smallskip}\hline
\end{tabular}
\end{table}
\paragraph{The resonance $2\dot g$}
Because of the orbital critical inclination, the lunisolar periodic component with argument $2g$ produces a non negligible contribution on the dynamics of the argument of the pericenter if compared with the oblateness one and with the precession due to the lunisolar mean terms.
Therefore, in the single resonance model of $2\dot g$ the asymmetry between the real equilibria yields that the ideal model SRM does not give reliable estimates of the resonant region, in accordance with \cite{cinesimoln}. Fig.~\ref{fig:2g} depicts the dynamics in $(e,2g)$ plane and in $(i,2g)$ plane around the main resonance. The maximum excursion in inclination, as computed with SRM, is $[58.03\hspace{0.1cm},67.36\hspace{0.1cm}\deg]$ and it is pretty similar to the one obtained with NSA in Tab.~\ref{tab:SRMno}, the difference being around one degree both for the minimum and the maximum inclination. The excursions in eccentricity given by the two models are quite different: the minimum value of the eccentricity reached in the libration region of the pendulum-like approximation is approximately $0.59$ and is quite different from $e_{min}=0.55$. The maximum values are both above the threshold of $e=0.76$: Molniya orbits with semi-major axis $\mathnormal{a}\approx\mathnormal{a}_{moln}$ cannot orbit with eccentricity larger than $0.76$ because the corresponding perigee would be smaller than the radius of the Earth. \\
\begin{figure*}
	\centering
	\begin{minipage}{0.35\textwidth}
		\includegraphics[width=0.8\textwidth]{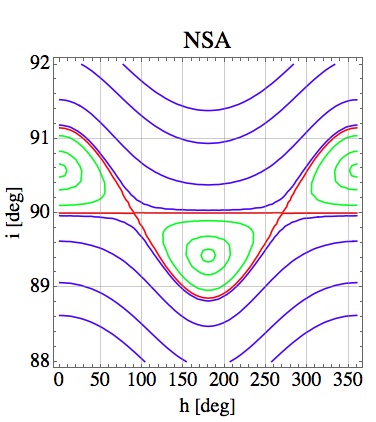}\\
	\end{minipage}%
	\caption{$\dot h$ resonance: contour plot of the resonant Hamiltonian. The X-axis shows the critical angle $h$ in $\deg$ and the Y-axis shows the $i$ in $\deg$.  \textit{Green} lines denote the librating curves around the stable equilibria, \textit{blue} lines denote the circulation region while the separatrices are sketched in \textit{red}.}
	\label{fig:h}       
\end{figure*}
\begin{figure*}
	\centering
	\begin{minipage}{0.35\textwidth}
		\includegraphics[width=0.8\textwidth]{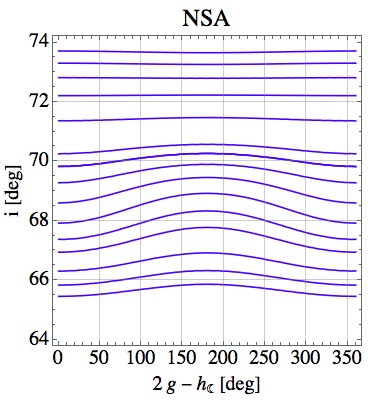}\\
	\end{minipage}%
	\begin{minipage}{0.35\textwidth}
		\includegraphics[width=0.8\textwidth]{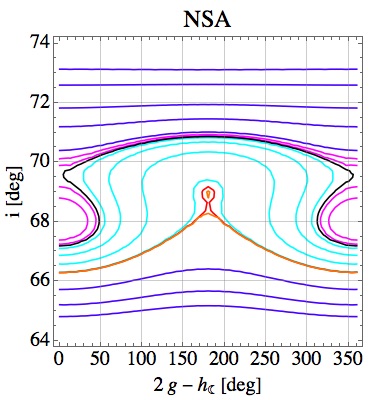}\\
	\end{minipage}%
	\begin{minipage}{0.35\textwidth}
		\includegraphics[width=0.8\textwidth]{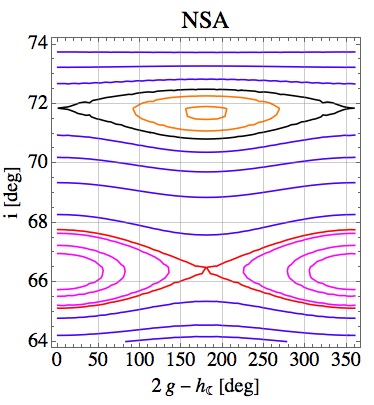}\\
	\end{minipage}%
	\caption{$2\dot g-\dot h_{\leftmoon}$ resonance: contour plot of the Hamiltonian obtained after performing the suitable coordinate change. The pictures show how the phase structure changes by using different initial condition to evaluate the first integral. $e=e_{moln}$ is the initial eccentricity, while the initial inclination varies: $i=64.3\hspace{0.1cm}\deg$ (on the left), $i=64.9\hspace{0.1cm}\deg$ (at the center), $i=66$ (on the right). \textit{Blue} lines identify the circulation orbits. The separatrices arising from the different unstable equilibria are drawn in \textit{red} and \textit{black} while the corresponding librating curves are denoted in \textit{magenta} and \textit{orange}, respectively. At the center, there is no clear distinction between the libration region associated with different stable equilibria. \textit{Cyan} lines represent the curves filling the overlapping region. }
	\label{fig:2gmhm} 
\end{figure*}
\paragraph{The resonance $\dot h$}
The polar resonance shows a non-standard behaviour around the inclination of $90\hspace{0.1cm}\deg$. As reported in Tab.~\ref{tab:SRMno} and depicted in Fig.~\ref{fig:h}, there are two stable equilibria with  $360\hspace{0.1cm}\deg$ of periodicity: the equilibrium on the prograde orbit region at $h=180$ and the one on the retrograde region at $h=0$. The corresponding unstable equilibria both lie at $i_u=90\hspace{0.1cm}\deg$ and the different libration regions result separated. In any case, the libration region does not overlap with the one of the resonances seen before. \\
\paragraph{The resonance $2\dot g -\dot{h}_{\leftmoon}$} By choosing different values of the first integral, in our case different initial eccentricity and inclination, the phase space structure drastically changes, as shown in the phase portraits in Fig.~\ref{fig:2gmhm}. In such case, the pendulum-like approximation is useless because of the bifurcation phenomenon.\\

Finally, by putting together the maximum and the minimum $i$ and $e$ that may be attained in the libration region of every single resonance we get the \textit{maximum overlapping region}: 
\begin{equation}\label{maxovreg}
e\in[0.44,0.79], \quad i\in [59.30\hspace{0.1cm}\deg,72.8\hspace{0.1cm}\deg]
\end{equation}
It is widely extended both in eccentricity and in inclination. This result could be the starting point for further investigation on the chaotic behaviour of Molniya orbits.

\section{Conclusion and discussion}
\begin{figure}
	\centering
	\includegraphics[width=0.8\textwidth]{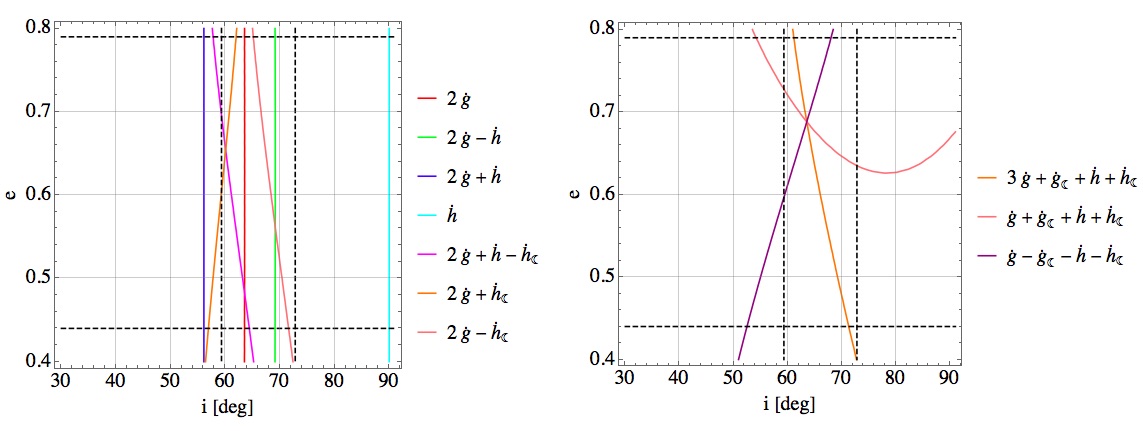}\\
	\caption{On the left, the second order resonances corresponding to the dominant terms taken into account are depicted in the $(e,i)$ plane. On the right the third order resonances with largest ratio (see Tab.~\ref{tab:ratio}) in $(e,i)$ plane. In both figures the dashed black horizontal lines lie in correspondence with the maximum and the minimum eccentricity reached in the region indicated in Eq. $\eqref{maxovreg}$, while, the dashed black vertical lines lie in correspondence with the maximum and the minimum inclination reached in the maximum overlapping region.}
	\label{fig:res3ord}       
\end{figure}

In this paper, the effects due to the lunisolar perturbation on the Molniya long-term dynamics have been studied with a rigorous analytical approach based on the Hamiltonian systems theory. We have built a doubly-averaged Hamiltonian including the oblateness secular effect and the lunisolar potential expansions up to the octupolar approximation. The perturbing contribution caused by each term appearing in the Hamiltonian model has been estimated by evaluating it with the Molniya parameters: the \textit{amplitude} and the \textit{period} in case of a periodic component or the precession/regression rate in case of a secular term accumulating over time. Using the Delaunay variables we noticed that the larger the \textit{amplitudes} the deeper the periodic fluctuation, while the periods help us to identify which harmonics produce long-term oscillations and which ones give rise to near-resonant or resonant terms. Finally, the results concerning the \textit{ratio} between amplitudes and the corresponding frequency confirm that the dynamics is governed by the second order lunisolar perturbation, as found numerically in \cite{molnarxiv}. In addition to the harmonics corresponding to $2g$ and $2g\pm h$, already taken into account in \cite{cinesimoln}, the long-term behaviour is strongly influenced also by perturbing terms associated with the argument $h$ and with some arguments involving the lunar ascending node.\\
The role of the third-body effect is crucial also for the evolution of the argument of the pericenter, the critical inclination makes such effect to be dominant. For this reason, the SRM provides an approximation too weak to properly describe the real dynamics in a neighborhood of the main resonance $2\dot g$. 
Furthermore, the ideal pendulum-like model fails both for the bifurcation phenomenon related to $2\dot g-\dot{h}_{\leftmoon}$ and for the non-standard behaviour around the polar resonance.
The identification of a maximum overlapping region could be a starting point for further investigation of the chaotic behaviour.\\
The third-order lunisolar perturbation does not seem to be particularly significant as regards to the dynamics, but, it could play a more important role in relation to chaotic phenomena. Only three third-order resonances show a \textit{ratio} larger than few second order terms. From Fig.~\ref{fig:res3ord}, which depicts the location of such resonances, we expect that $3g+g_{\leftmoon}+h+h_{\leftmoon}$, $g+g_{\leftmoon}+h+h_{\leftmoon}$ and $g-g_{\leftmoon}-h-h_{\leftmoon}$ overlap with the maximum overlapping region found in Eq. $\eqref{maxovreg}$. In any case, if no \textit{anomalous} dynamical behaviour occurs, such as bifurcations, we may expect that the third-order resonances show a quite narrow libration region for which the SRM provides a well-approximation. 

\end{document}